\pdfoutput=0
\documentclass[twoside,reqno,a4paper,12pt]{scrartcl}

\usepackage{psfig}
\usepackage{amsmath,amsmath, amssymb}
\usepackage{epsfig}

\usepackage[authoryear]{natbib}
\RequirePackage{hyperref}
\RequirePackage{hypernat}

\renewcommand{\baselinestretch}{1.5}
\usepackage{scrpage2}

\setkomafont{section}{\bf \rmfamily}
\setkomafont{subsection}{\bfseries \itshape \rmfamily}
\clearscrheadfoot
\chardef\at=`\@
\DeclareRobustCommand{\qed}{%
  \ifmmode 
  \else \leavevmode\unskip\penalty9999 \hbox{}\nobreak\hfill
  \fi
  \quad\hbox{\qedsymbol}}
\newcommand{\rz}{\mathbb R}

\newcommand{\pr}{\mathbb P}

\newcommand{\ex}{\mathbb E}
\newcommand{\mathbold}[1]{\mbox{\boldmath $#1$}}

\newtheorem{theorem}{Theorem}[section]

\newcommand{\eps}{\varepsilon}

\title{\bf \large Approximating Data with weighted smoothing Splines} 
\author{\large P.L.~Davies$^{a,b}$ and M.~Meise$^{a,} \footnote{Corresponding author. Email: monika.meise@uni-due.de} $\bigskip\\{\large \em $^a$University of Duisburg-Essen, Germany;}\\{\large \em $^b$TU Eindhoven, Netherlands}}
\date{}
\begin{document}
\pagestyle{scrheadings}
\maketitle

\begin{abstract}
\begin{small}
Given a data set $(t_i, y_i)$, $i=1, \dots, n$ with the $t_i \in
[0,\,1]$ non-parametric regression is concerned with the 
problem of specifying a suitable function $f_n:[0,\,1] \rightarrow
\rz$ such that the data can be reasonably approximated by the points
$(t_i, f_n(t_i))$, $i=1, \dots, n.$  If a data set exhibits large
variations in local behaviour, for example large peaks as in
spectroscopy data, then the method must be able to adapt to the local
changes in smoothness. Whilst many methods are able to accomplish this
they are less successful at adapting derivatives. In this paper we
show how the goal of local adaptivity of the function and its first
and second derivatives can be attained in a simple manner using
weighted smoothing splines. A residual based concept of approximation
is used which forces local adaptivity of the regression function
together with a global regularization which makes the function as
smooth as possible subject to the approximation constraints.\\
{\em AMS 2000 Subject classifications:} Primary 62G08, secondary 62G15, 62G20\\
{\bf Keywords:} nonparametric regression, smoothing splines, confidence region, regularization
\end{small}
\end{abstract}

\newcounter{fig1}
\newcounter{fig2}

\section{Introduction} 
\subsection{Smoothing and weighted smoothing splines}
\begin{figure}
  \centering
  \psfig{file=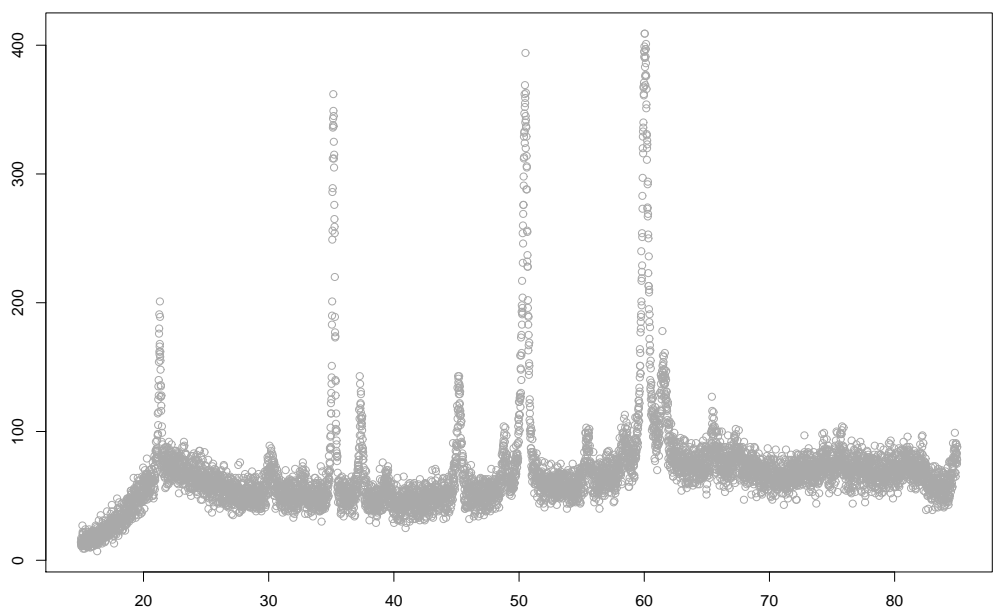,angle=270,width=14cm}
  \caption{ {\footnotesize Data from thin-film physics showing the
      photon count of X-rays as a function of the angle of diffraction
      measured in degrees.}} \label{FIGm35020a}
\end{figure}

In the one-dimensional case nonparametric regression is concerned
with determining a function $f_n: [0,\,1] \rightarrow \rz$ which
adequately represents a data set ${\mathbold y}_n=\{(t_i,y(t_i)): t_i \in
[0,\,1],\,i=1,\ldots,n\}.$  The problem is to provide a function
$f_n$ which is an adequate representation of the data. One well
established method for accomplishing this goal is that of smoothing
splines defined as the solution of the problem
\begin{eqnarray} \label{eq:SS}
 \min S(g,\lambda):= \sum_{i=1}^n (y(t_i)-g(t_i))^2+\lambda
 \int_0^1g^{(2)}(t)^2\,dt  
\end{eqnarray}
where $\lambda$ is the smoothing parameter (see \cite{Wahba1990,
  GRESILV94,RUPWANCAR03}).  
This approach has two weaknesses. The first
is that there may not be any choice of $\lambda$ for which the
resulting fit is satisfactory. This is particularly the
case if the data show large local variations such as in Figure
\ref{FIGm35020a} which are taken from thin film physics. They were  kindly
supplied by Prof.~Dieter Mergel of the Department of Physics, 
University of Duisburg-Essen. X-rays are beamed onto a thin film and
the data give the photon count of the diffracted rays as a function of
the angle of diffraction. The sample size is $n=7001.$ The high peaks
can only be adequately captured with a small value of $\lambda$ in
(\ref{eq:SS}). This has however the consequence that the function
oscillates too rapidly between the peaks. The second problem is to
give an automatic choice for $\lambda.$ Methods suggested include
cross-validation, generalized cross-validation, generalized maximum
likelihood and restricted maximum likelihood (\cite{CRAWAH78,
  Wahba1985, RUPWANCAR03}). However it is clear that if there is no
satisfactory value of $\lambda$ then no automatic choice will work.

In this paper we attain more flexibility by considering a vector
${\mathbold \lambda}=(\lambda_1,\ldots,\lambda_n)$ rather than a
single value $\lambda$ and we replace the minimization
problem (\ref{eq:SS}) by
\begin{eqnarray} \label{eq:WSS}
 \min S(g,{\mathbold\lambda}):=
\sum_{i=1}^n\lambda_i(y(t_i)-g(t_i))^2+\int_0^1g^{(2)}(t)^2\,dt.   
\end{eqnarray}
Comparing this with  (\ref{eq:SS}) we see that the smoothing
parameter $\lambda$ has now been transferred from the penalty term to
the observations themselves. The solution, which we denote by
$f_n(\cdot:{\mathbold \lambda}),$ is a natural cubic spline (see
\cite{GRESILV94}) but the $\lambda_i$ now control the fit at the
observation points $(t_i,y(t_i))$ rather than the size of the penalty
which is now fixed. In the case of the data displayed in Figure
\ref{FIGm35020a} we would choose large values of $\lambda_i$ at the
peaks causing them to be adequately approximated. At points away from
the peaks we would choose the $\lambda_i$ to be small and thus ensure
a smooth solution at these points.

The method proposed here belongs to the category of spatially adaptive
splines. For other spatially adaptive spline methods we refer to
 \cite{LuoWahba1997}, \cite{DEMASM98}, \cite{RUPPCARR00}, \cite{ZHSH01},
\cite{DimatGenKass01}, \cite{PITT02}, \cite{WOJITA02}, \cite{MISH03,MISH05}, \cite{PISPEHO06}.

\subsection{Contents}
In Section 2 we describe an approach to choosing a model in the
context of nonparametric regression which is based on a universal,
honest and non-asymptotic confidence region. Section 3 shows how the ideas
of Section 2 can be adapted to give a simple method for choosing the
weights of a weighted smoothing spline. Examples and the results of a
small simulation study are given in Section 4. Section 5 gives two
variations on this theme and Section 6 extends the method to image
analysis. Finally in Section 7 we look at the asymptotics. 

\section{Choosing a model}
\subsection{Nonparametric confidence regions}
A lot of work has been devoted to choosing a model from a
sequence of models of increasing complexity. Choosing a value of
$\lambda$ in (\ref{eq:SS}) falls into this category as the smaller
$\lambda$ the more complex the resulting smoothing spline. Methods
developed to solve the problem include cross-validation, plug-in
methods as well as AIC and BIC which are explicitly phrased in terms
of balancing complexity and fidelity. We take a different approach
here which is implicit in \cite{DAVKOV04} and explicit in
\cite{DAVKOVMEI08}. We define a universal, honest and non-asymptotic
confidence region ${\mathcal A}_n$ and given this region we choose
non-decreasing $\lambda_i^j,\,j=1,2,\ldots$  to force
$f_n(\cdot:{\mathbold \lambda}^j)$ to eventually lie in ${\mathcal
  A}_n.$ This gives a sequence of functions of increasing
roughness (or complexity) and we choose the first one which lies in
${\mathcal A}_n.$ The region ${\mathcal A}_n$ is based on the
residuals and requires a stochastic model. The one we use is 
\begin{eqnarray} \label{basicmod}
Y(t)=f(t)+\sigma Z(t), \quad 0 \le t \le 1
\end{eqnarray}
with $Z(t)$ standard Gaussian white noise. Following
\cite{DAVKOVMEI08} ${\mathcal A}_n$ is defined as 
follows. For any function $g$ we consider normalized sums of residuals
over intervals 
\begin{eqnarray} \label{defwifn}
w({\mathbold y}_n,I,g)=\frac{1}{\sqrt{\vert I \vert}}\,\sum_{t_i\in
  I}(y(t_i)-g(t_i))
\end{eqnarray}
where $\vert I\vert$ denotes the number of points $t_i$ in the
interval $I.$ For data ${\mathbold Y}_n={\mathbold Y}_n(f)$ generated
under the model (\ref{basicmod}) we define the confidence region for
$f$ by
\begin{equation*} \label{conreg1}
{\mathcal A}_n= {\mathcal A}_n({\mathbold Y}_n, \sigma,{\mathcal
  I}_n,\tau_n)=\big\{g: \max_{I \in {\mathcal  I}_n} \,\vert
w({\mathbold Y}_n,I,g)\vert \le \sigma\sqrt{\tau_n \log n\,}\,\big\}
\end{equation*}
where ${\mathcal I}_n$ is a family of intervals and
$\tau_n=\tau_n(\alpha)$ is defined by
\begin{eqnarray} \label{sizetest}
P\Big(\max_{I\in {\mathcal I}_n}\frac{1}{\sqrt{\vert I\vert}}
\Big\vert \sum_{i\in   I}Z(t_i)\Big\vert \le \sqrt{\tau_n\log n\,}\,
\Big)=\alpha. 
\end{eqnarray}
It follows that ${\mathcal A}_n({\mathbold Y}_n,
\sigma,{\mathcal I}_n,\tau_n)$ is  a universal, honest and
non-asymptotic confidence region  for $f$, that is
\begin{equation*} \label{exuninasym}
{\mathbold P}(f \in {\mathcal A}_n({\mathbold Y}_n(f),\sigma,{\mathcal
  I}_n, \tau_n))=\alpha \quad \text{for all}\,\,f\,\,\text{and}\,\,n.
\end{equation*}
The family of intervals ${\mathcal I}_n$ can be taken to be
the family of all intervals but this is computationally
expensive. For all practical purposes it suffices to consider a subfamily
of intervals as long as it is multiscale, that is, if it contains
intervals of all sizes. The simplest such scheme, and the one we shall
use, corresponds closely to that defined by the Haar wavelet. If
$n=2^m$ the family ${\mathcal I}_n$ consists of all
one-point intervals $[t_1,t_1] \ldots,[t_n,t_n]$, all two point intervals
$[t_1,t_2],[t_3,t_4],\ldots, [t_{n-1},t_n]$, all four-point
intervals $[t_1,t_4],[t_5,t_8],\ldots,[t_{n-3},t_n]$ and so forth. If
$n$ is not a power of 2 we simply include the last interval 
whatever its form. In the remainder of the paper we use this dyadic
scheme. For any scheme ${\mathcal I}_n$ and for given $\alpha$ the
values of $\tau_n(\alpha)$ as defined by (\ref{sizetest}) can be
obtained by simulations.  Table \ref{tauval} gives the values of
$\tau_n(\alpha)$ for the dyadic scheme just described, $\alpha=0.95$
and 0.99  and for various sample sizes $n.$ The results are based on
10000 simulations.  

\begin{table}
\begin{center}
\begin{tabular}{|c|ccccccc|}
\hline
$n$&100&250&500&1000&2500&5000&10000\\
\hline
0.95&2.92&2.88&2.79&2.71&2.64&2.60&2.55\\
0.99&3.60&3.41&3.33&3.17&3.03&3.00&2.92\\
\hline
\end{tabular}
\caption{{\footnotesize The values of $\tau_n(\alpha)$ for the dyadic
    scheme ${\mathcal I}_n,\,\alpha=0.95$ and 0.99 and
    $n=100,\,250,\,500,\,1000,\,2500,\,5000$ and 10000.}\label{tauval}} 
\end{center}
\end{table}
 It follows from a result of \cite{DUEMSPO01} and the very precise
 result of \cite{KAB07} that if ${\mathcal I}_n$ contains all
 one-point  intervals then  
\[ \lim_{n \rightarrow \infty} \tau_n(\alpha)=2\]
for all $\alpha.$ In particular this holds for the dyadic multiscale
family ${\mathcal I}_n$ we consider. The resulting curves are not
sensitive to the value of $\tau_n(\alpha)$ and so for simplicity in
the remainder of the paper we simply put $\tau_n(\alpha)=3.$ This is
consistent with the values of Table \ref{tauval}. 

\begin{figure}
  \centering
  \psfig{file=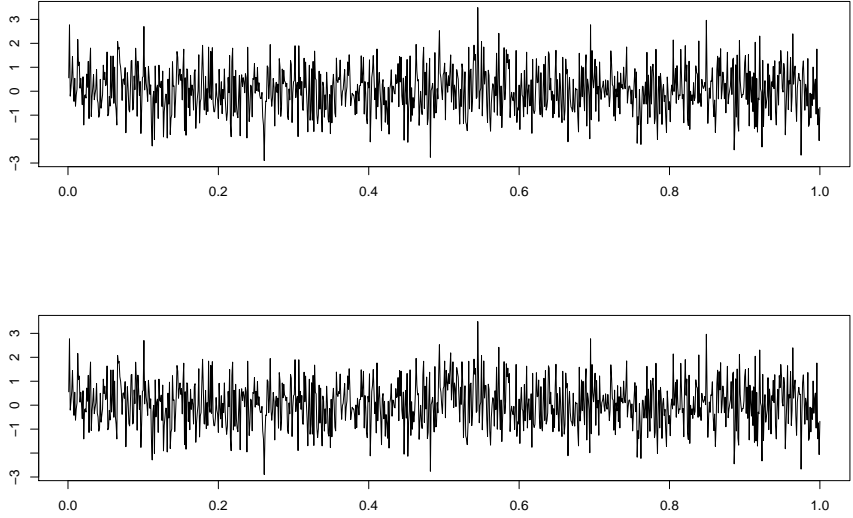,angle=270,width=12cm}
  \caption{The upper panel shows standard white noise $Z(t)$. The
    lower panel shows $f'(t)+Z(t)$ with ${\tilde f}_n(t)=0.7$ for $0.5 <
    t \le 0.524$ and zero otherwise.} 
\label{suffall} 
\end{figure}

An Associate Editor asked to what extent the results depend on the
chosen scheme ${\mathcal I}_n$ and the value of $\tau_n.$ This can be
analysed as follows. Suppose the data are generated by a function $f$
and consider a function ${\tilde f}_n$ which differs from $f$ by
$\delta_n$ on an interval $I$, that is ${\tilde f}_n(t)-f(t) >\delta_n,\,
t \in I.$ This will be detected by the procedure if ${\tilde f}_n \notin
{\mathcal A}_n.$ If ${\mathcal I}_n$ is the family of all intervals then
${\tilde f}_n \notin {\mathcal A}_n$ follows from
$$\frac{1}{\sqrt{\vert I\vert}}\sum_{t_i \in I} (Y(t_j)-{\tilde
  f}(t_j)) \le   -\sigma\sqrt{\tau_n \log n}.$$
From this we deduce that the deviation will be detected with
probability at least $\alpha-0.01$ if 
\begin{equation} \label{detectdiff1}
\delta_n \ge \sigma\Big(\sqrt{\tau_n\log n}+ 2.3263\Big)/\sqrt{\vert
  I\vert}\,.
\end{equation}   
If we use the dyadic scheme ${\mathcal I}_n'$ it is no longer
guaranteed that $I \in {\mathcal I}_n'.$ However there exists an
interval $I'\subset I$ in ${\mathcal I}_n'$ with $\vert I'\vert \ge
\vert I\vert/2.$ The same argument gives
\begin{equation} \label{detectdiff2}
\delta_n \ge \sigma\sqrt{2}\Big(\sqrt{\tau_n'\log n}+
2.3263\Big)/\sqrt{\vert I\vert}\,
\end{equation}   
Denser schemes ${\mathcal I}_n(\kappa)$ parameterized by a parameter
$\kappa, \,1 < \kappa \le 2,$ with $\vert{\mathcal I}_n(\kappa) \vert
=O(n)$ are given in \cite{DAVKOVMEI08}: the dyadic scheme corresponds
to the case $\kappa=2$. If we use ${\mathcal I}_n(\kappa)$ then we
can replace (\ref{detectdiff2}) by 
\begin{equation*} \label{detectdiff3}
\delta_n \ge \sigma\sqrt{\kappa}\Big(\sqrt{\tau_n(\kappa)\log n}+
2.3263\Big)/\sqrt{\vert I\vert}\,.
\end{equation*}   
As $\tau_n(\kappa) < \tau_n$ this can be made arbitrarily close to the
case of all intervals (\ref{detectdiff1}). The dyadic scheme is the
coarsest we use, but it is nevertheless efficacious as shown by the
results of \cite{DAVGATWEI08}. The analysis we have done is for a
worst-case situation, the actual performance may be better. As an
example we take $\alpha=0.95,\, \sigma=1$ and $n=1000.$ It follows
from Table \ref{tauval} that the value of $\tau_n'$ in
(\ref{detectdiff2}) is 2.71. Simulations show that the corresponding
value of $\tau_n$ in (\ref{detectdiff1}) is 2.91. If $\vert I\vert=24$
then we have $\delta_n \ge 1.39$ for (\ref{detectdiff1}) and $\delta_n
\ge 1.92$ for (\ref{detectdiff2}). The upper panel of Figure
\ref{suffall} shows standard white  noise $f \equiv 0$: the lower
panel shows ${\tilde f}_n(t)+Z(t)$ for the same noise with ${\tilde
  f}(t)=\delta$ for $0.5<t \le 0.524,$ zero otherwise and $\delta=0.7.$
The signal in the lower panel is difficult to detect by eye: the
signal-to-noise ratio is 0.11. However it is detected using the
dyadic scheme with $\tau_n=3$. If we put $\tau_n=2.71$ then a signal
with $\delta=0.63$ is detected. If we use all intervals then with
$\tau_n=3$ a signal with $\delta=0.67$ is detected, for $\tau_n=2.91$
a signal with $\delta=0.64$ is detected. The differences are not large.

So far we have assumed that $\sigma$ is known which is not the
case. We use the default value (\cite{DAVKOV01})
\begin{eqnarray} \label{sigma1}
\sigma_{n}= \frac{1.4826}{\sqrt{2}}\text{MED}\{\vert
y(t_i)-y(t_{i-1})\vert, i=2,\dots,n-1\}. 
\end{eqnarray}
For data generated under the model we have
\[Y(t_i)-Y(t_{i-1})=Z(t_i)-Z(t_{i-1})+f(t_i)-f(t_{i-1}).\]
If $Z$ is a $N(0,1)$ random variable it may be checked that the median
of $\vert Z -c\vert$ strictly exceeds that of $\vert Z\vert$ for any $c\ne 0.$
From this it follows that $\sigma_n$ is always biased upwards under 
the model. Consequently ${\mathcal A}_n({\mathbold
  Y}_n,\sigma_n,{\mathcal I}_n,\tau_n)$ is no longer a universal,
exact, non-asymptotic confidence region 
but it is a universal,  honest (\cite{LI89}), non-asymptotic
confidence region   
\begin{equation*} \label{honuninasym}
{\mathbold P}( f \in  {\mathcal A}_n({\mathbold
  Y}_n,\sigma_n,{\mathcal I}_n,\tau_n)) \ge \alpha  \quad \text{for
  all}\,\,f\,\,\text{and}\,\,n. 
\end{equation*}

Given the confidence region ${\mathcal A}_n({\mathbold
  y}_n,\sigma_n,{\mathcal I}_n,\tau_n)$ and the measure of roughness
\begin{equation*} \label{measrough}
R(g)=\int_0^1 g^{(2)}(t)^2\,dt
\end{equation*} 
the natural approach would be to solve
\begin{equation} \label{minrough}
\text{minimize}\quad R(g) \quad\text{subject to}\quad
g \in {\mathcal A}_n({\mathbold
  y}_n,\sigma_n,{\mathcal I}_n,\tau_n).
\end{equation}
As ${\mathcal A}_n$ is defined by a set of linear inequalities
involving the values of $g$ at the points $t_i$  the problem is one of
quadratic programming. If we take the dyadic scheme for ${\mathcal
  I}_n$ then ${\mathcal A}_n$ is defined by about $4n$ linear
inequalities. For small data sets with $n \le 1000$ which 
exhibit little local variability it is possible to solve this directly
but the approach fails for data sets such as those of Figure
\ref{FIGm35020a} with $n=7001.$ The quadratic programming problem
involves 7001 parameters and the number of linear constraints is about
28000. Furthermore the fact that the squared second derivative varies
by several orders of magnitude over the interval causes excessive
numerical instability. In contrast the problem (\ref{eq:WSS}) can be
solved for in a fast and stable manner even for values of $\lambda_i$
which differ by orders of magnitude. In the next 
section we describe an automatic procedure for doing this which
attempts to emulate the solution of (\ref{minrough}).

The idea of the confidence region as defined above is implicit in
\cite{DAVKOV01}. A similar idea was used by
D\"umbgen and Spokoiny (2001) for testing for monotonicity and convexity of
nonparametric functions.  Universal, exact, non-asymptotic
confidence regions based on the signs of the residuals
$\text{sign}(y(t_i)-g(t_i))$ rather than the residuals themselves are
to be found implicitly in \cite{DAV95} and explicitly
in \cite{DUEM03, DUEM07} and \cite{DUEMJO04}. These require only that
under the model the errors are independently distributed with median
zero. As a consequence they do  not require an auxiliary estimate of
scale such as (\ref{sigma1}).

\section{Choosing the weights} 
\subsection{The procedure \label{wssproc}}
The procedure we use is based on the following heuristic. If $\Vert
{\mathbold \lambda}\Vert$ is small then the solution
$f_n(\cdot:{\mathbold \lambda})$ of (\ref{eq:WSS})
will be essentially the least squares line through the data. If on the
other hand all the components $\lambda_i$ of ${\mathbold \lambda}$ are
very large then $f_n(\cdot:{\mathbold \lambda})$ will almost
interpolate that data and will lie in ${\mathcal A}_n$ as
all residuals will be close to zero. The idea is then to start with
very small $\lambda_i$ and then to increase them gradually until
$f_n(\cdot:{\mathbold \lambda})$ lies
in ${\mathcal A}_n$ and then stop. More formally we start with the
least squares regression line and check whether this lies in
${\mathcal A}_n$. If so we stop and  accept the solution. Otherwise put
${\mathbold \lambda}^1=(\lambda_1,\ldots,\lambda_1)$
where $\lambda_1$ is chosen to be so small that the solution of
(\ref{eq:WSS}) with ${\mathbold \lambda}={\mathbold \lambda}^1$
differs from the least squares lines by some small prescribed
quantity. At the $i$th stage we have the solution
$f_n(\cdot:{\mathbold \lambda}^i)$ based on the weights ${\mathbold
  \lambda}^i.$ We check if the solution lies in ${\mathcal A}_n$ and
if so we stop. If not we determine those intervals $I_i \in {\mathcal
  I}_n$ for which
\begin{equation} \label{badint}
w({\mathbold y}_n,I_i,f_n(\cdot:{\mathbold \lambda}^i))\ge
\sigma_n\sqrt{\tau_n\log n}.
\end{equation}
For all points $t_j$ in any such interval we increase the
corresponding $\lambda_j^i$ by a factor of $q$, that is
$\lambda_j^{i+1}=q\lambda_j^i.$  Our default value for $q$ is 2. The
remaining $\lambda_j^i$ are not altered. This gives us a new
${\mathbold \lambda}^{i+1}$ and we repeat the procedure.  

As defined the procedure is difficult to analyse, especially as the
effect is a finite sample one: it will gradually disappear for a fixed
function $f$ as the sample size $n$ tends to infinity. The
problem can be circumnavigated to a certain extent as follows. We
consider a second procedure but this time with the components
$\lambda^i_j$ of ${\mathbold \lambda}^i$ all equal,  $\lambda^i_j=\lambda^i,
j=1,\ldots,n.$ If the solution does not lie in ${\mathcal A}_n$ then
all components are increased by a factor of $q$ and not just those
whose $t_j$ values lie in intervals $I_i$ for which (\ref{badint})
holds. For this form of ${\mathbold \lambda}=(\lambda,\ldots,\lambda)$
it can be shown that $R( f_n(\cdot:{\mathbold \lambda}))$ depends
monotonically on $\lambda$ which makes it amenable to mathematical
analysis. If we now perform both procedures and then choose at the end
the smoothest of the two solutions we have a procedure which can be
analysed. We have not yet encountered a data set where the result of the
second procedure with equal weights was chosen.
We point out that solving (\ref{eq:WSS}) for this form of ${\lambda}$
is equivalent to solving (\ref{eq:SS}) but with $\lambda^{-1}$ in
place of $\lambda.$ The second procedure therefore does the
following. It considers the one-dimensional family of solutions of
(\ref{eq:SS}) and chooses the smoothest such function which lies in
${\mathcal A}_n.$ This is an alternative to choosing the smoothing
parameter by cross-validation or likelihood methods. 

\subsection{An illustration}
\begin{figure}
  \centering
  \psfig{file=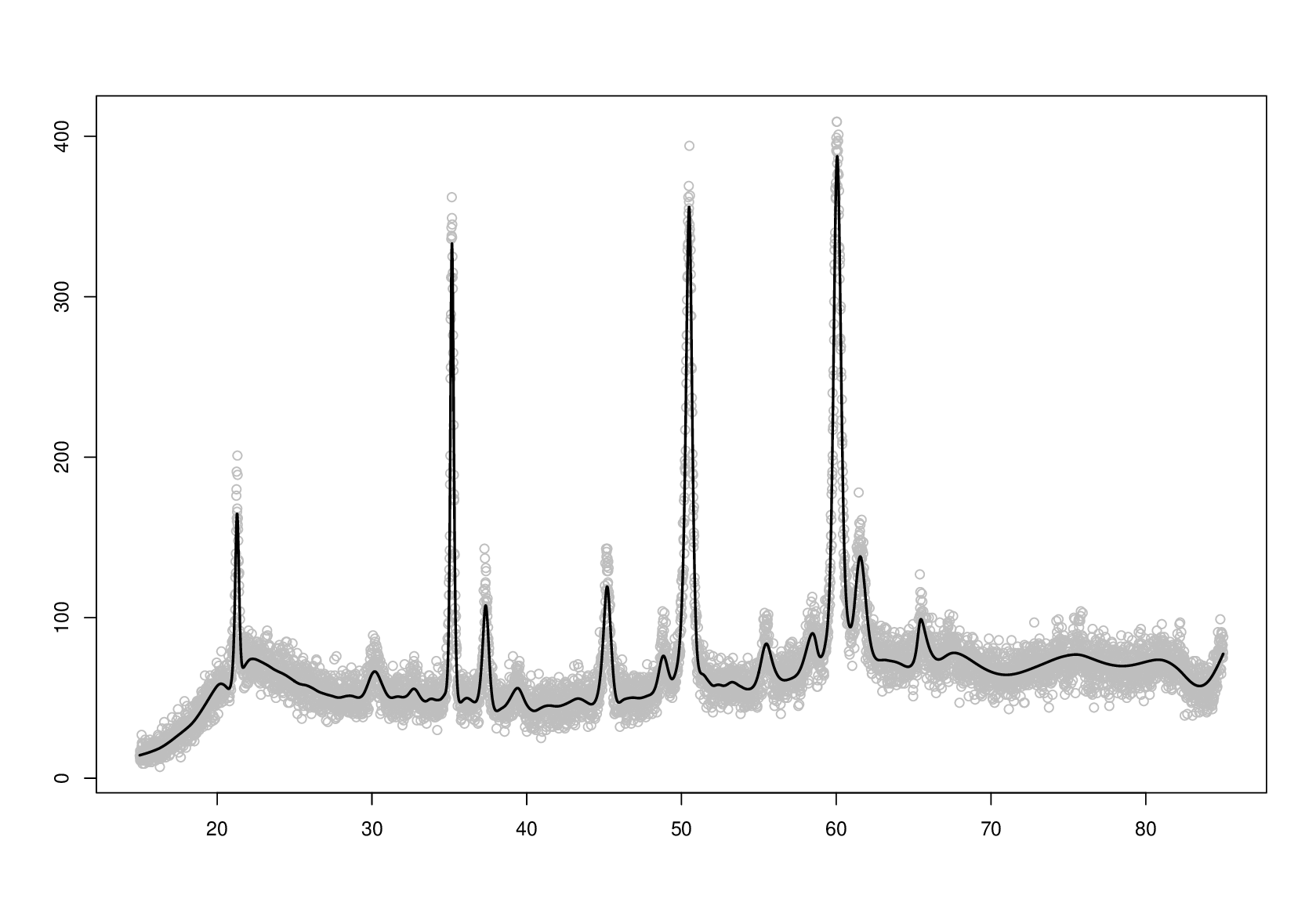,height=14cm,width=7cm,angle=270}\\
  \psfig{file=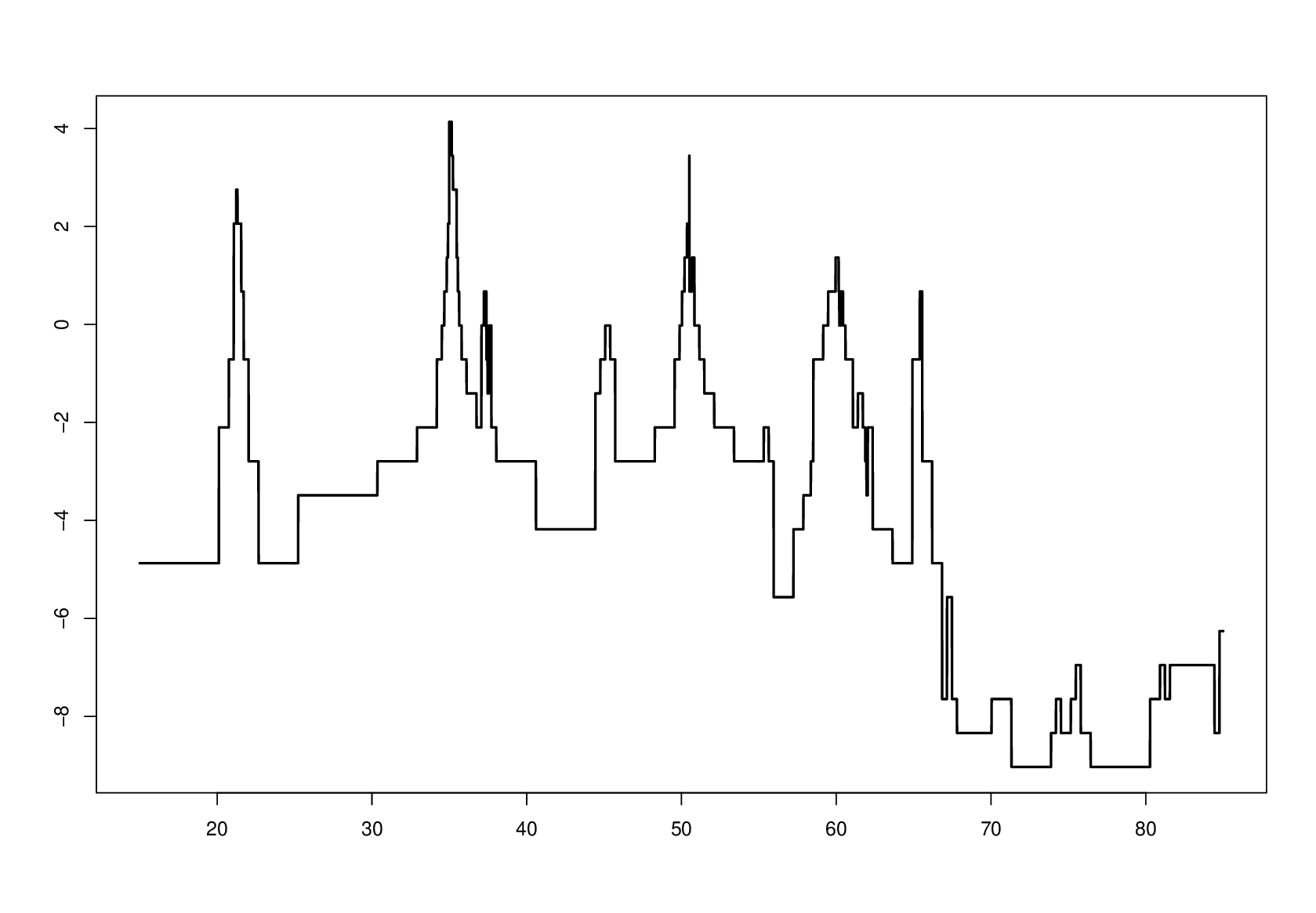,height=14cm,width=7cm,angle=270}
   \caption{{\footnotesize The upper panel shows the results of
       applying the weighted smoothing spline procedure to the thin
       film data. The lower panel shows the values of the $\lambda_i$
       on a logarithmic scale.} \label{wssthinfilmw}} 
\end{figure}

\begin{figure}
  \centering
  \psfig{file=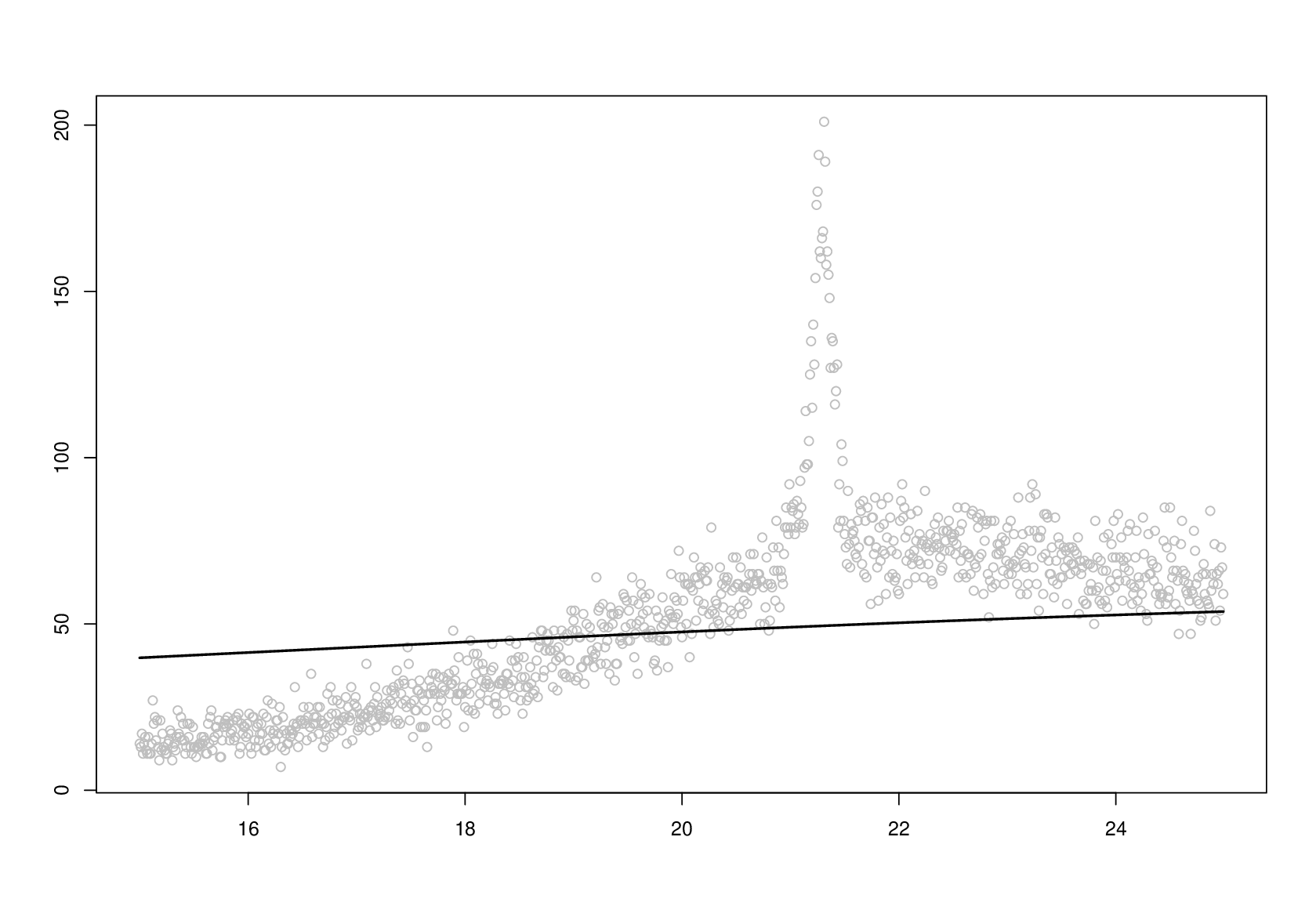,angle=270,width=7cm}
  \psfig{file=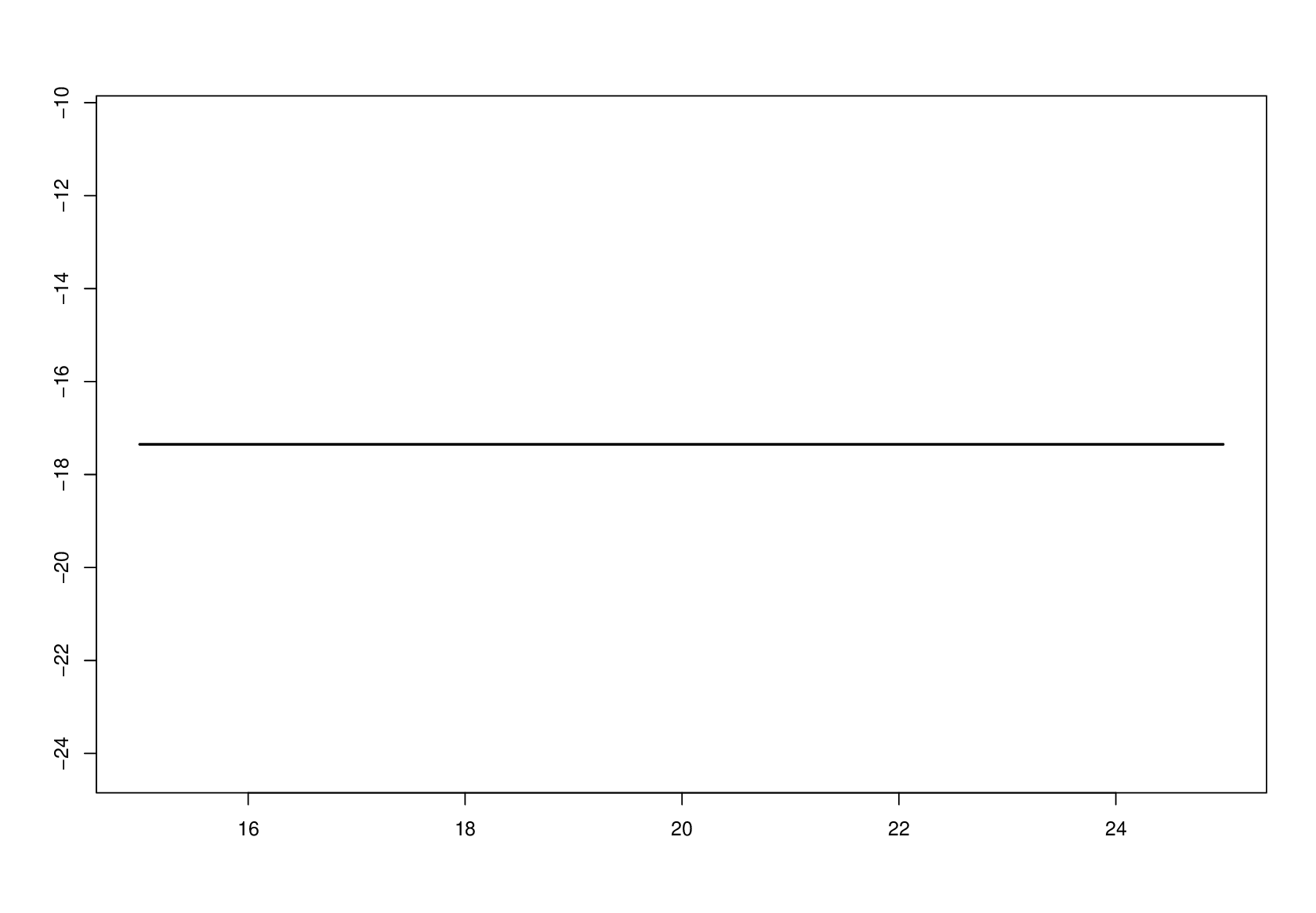,angle=270,width=7cm}\\
  \psfig{file=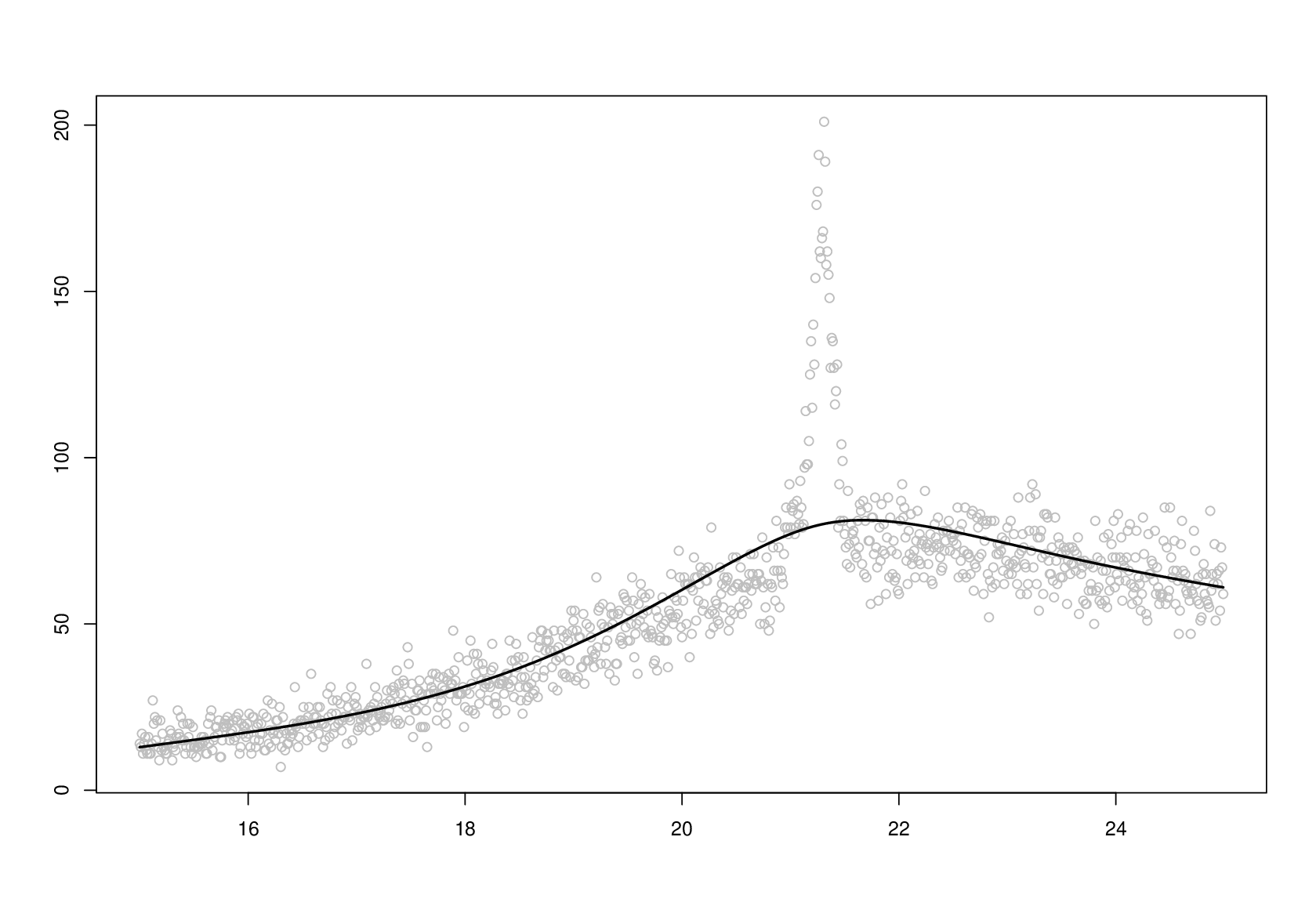,angle=270,width=7cm}
  \psfig{file=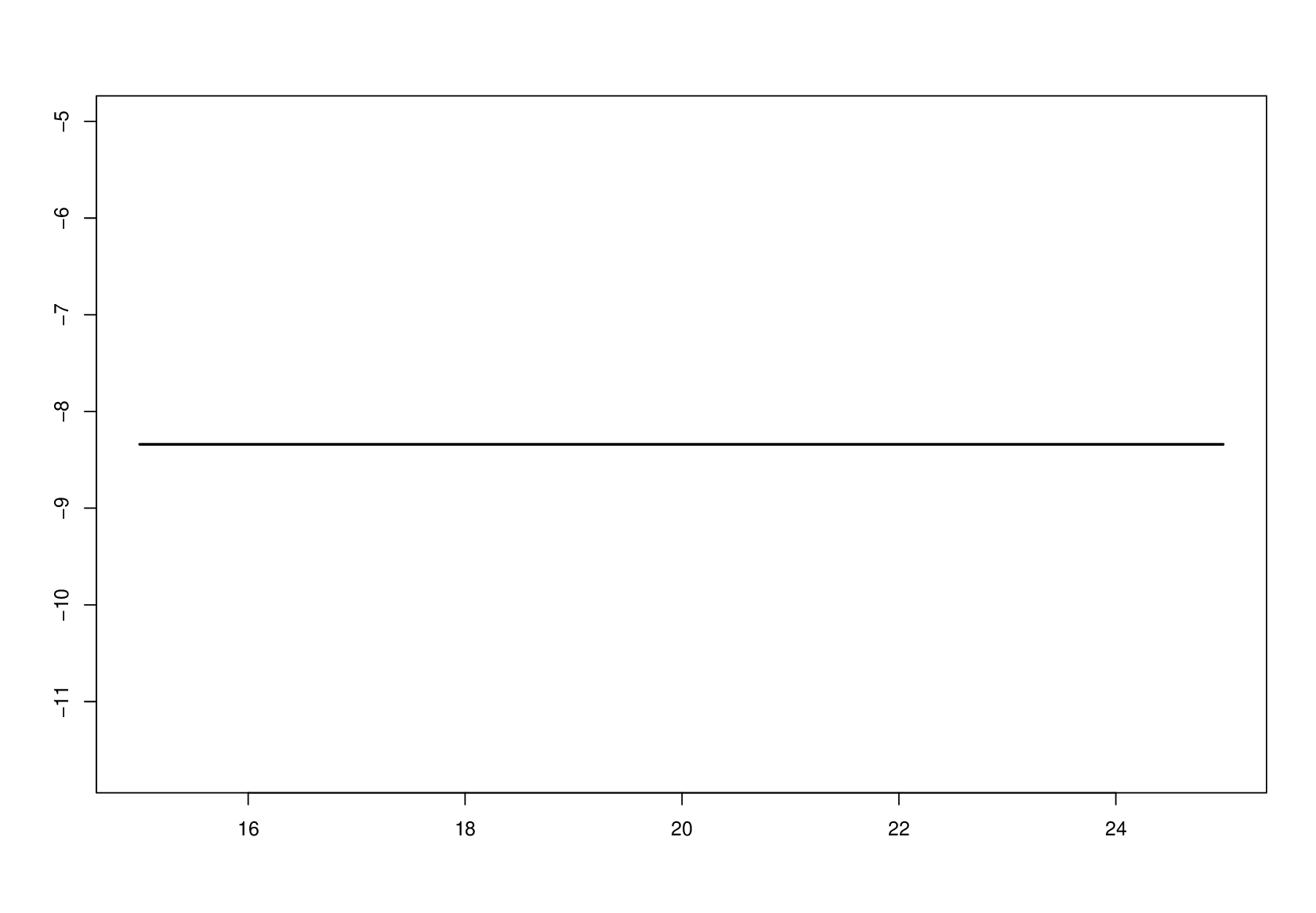,angle=270,width=7cm}\\
  \psfig{file=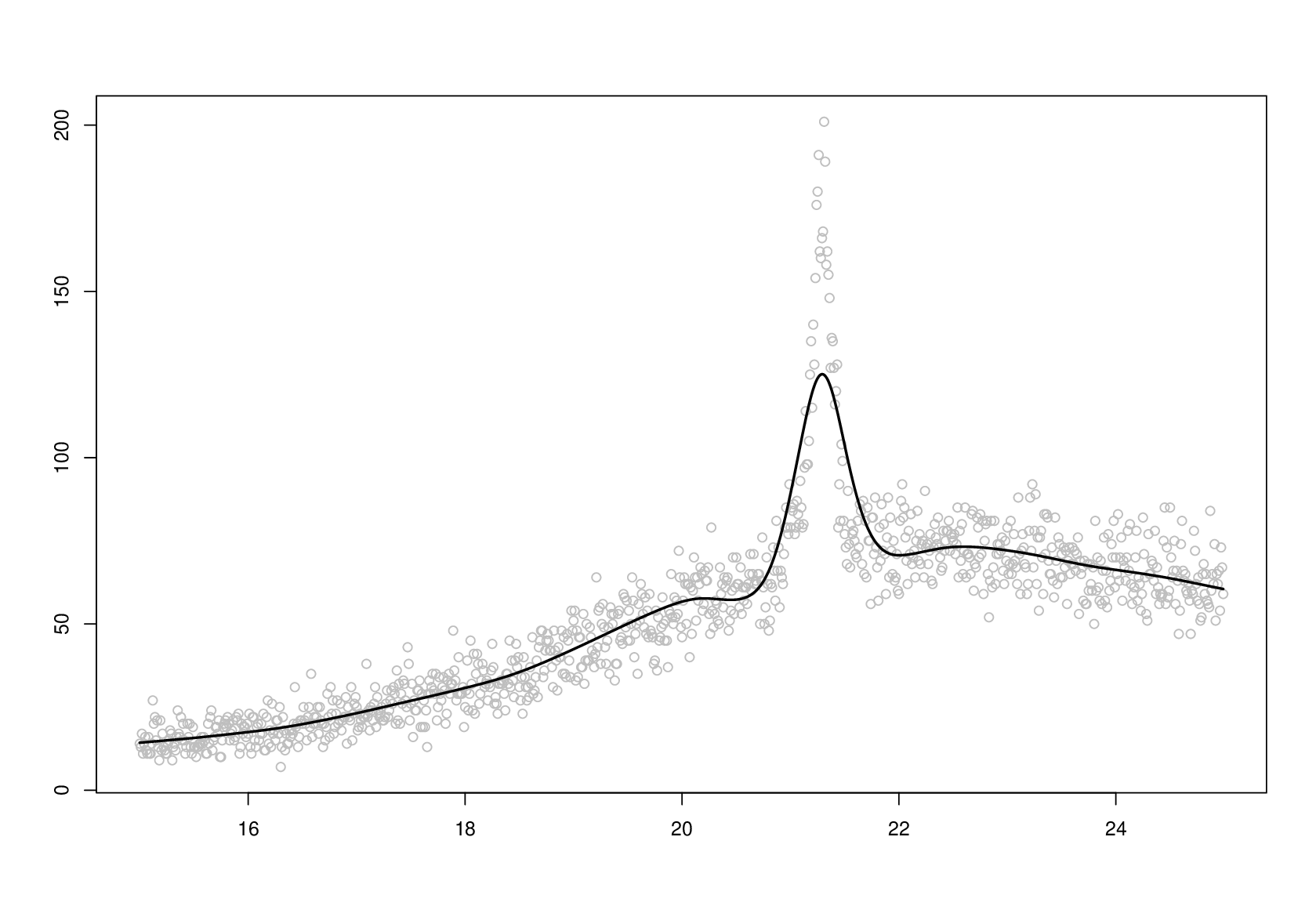,angle=270,width=7cm}
  \psfig{file=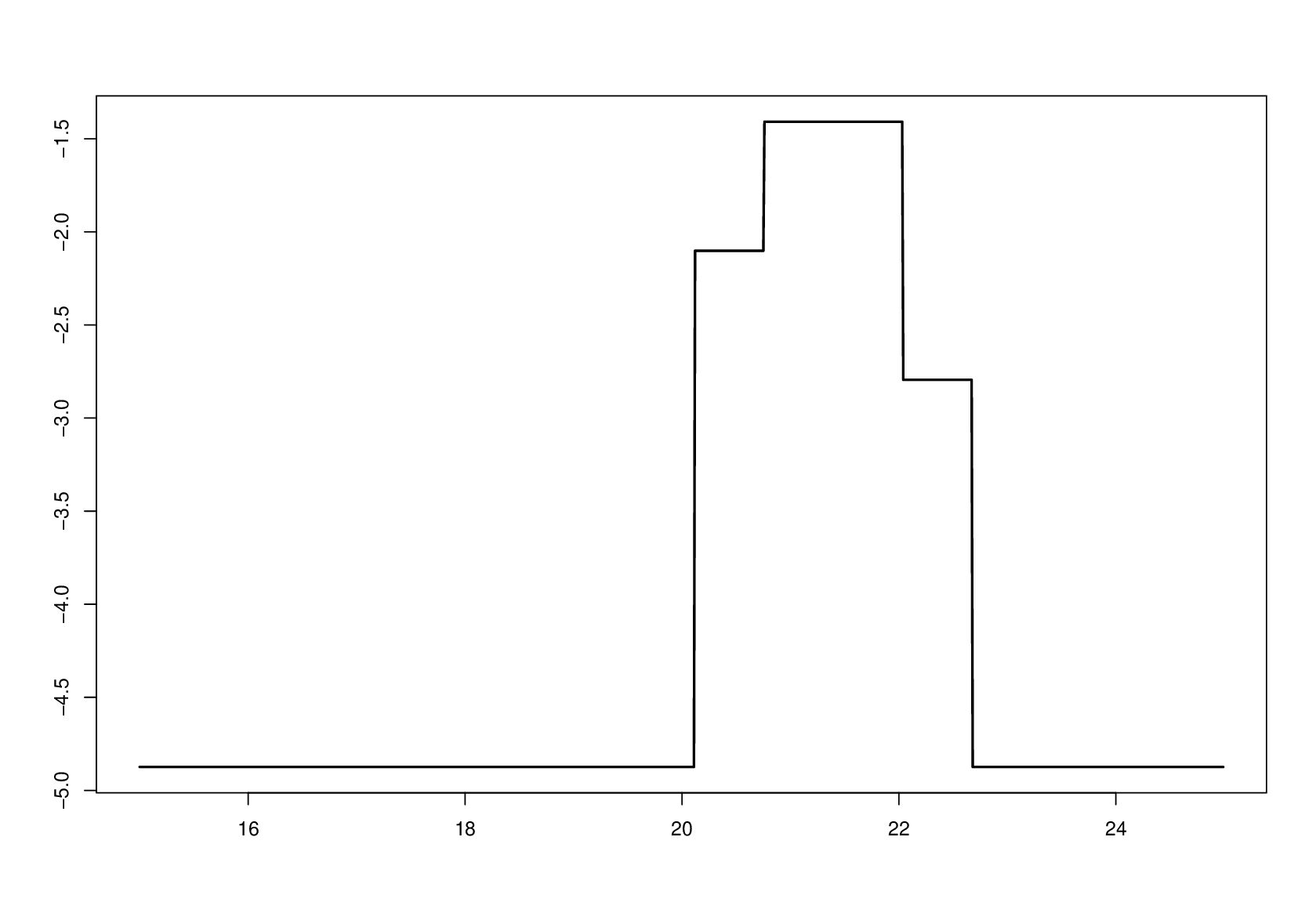,angle=270,width=7cm}\\
  \psfig{file=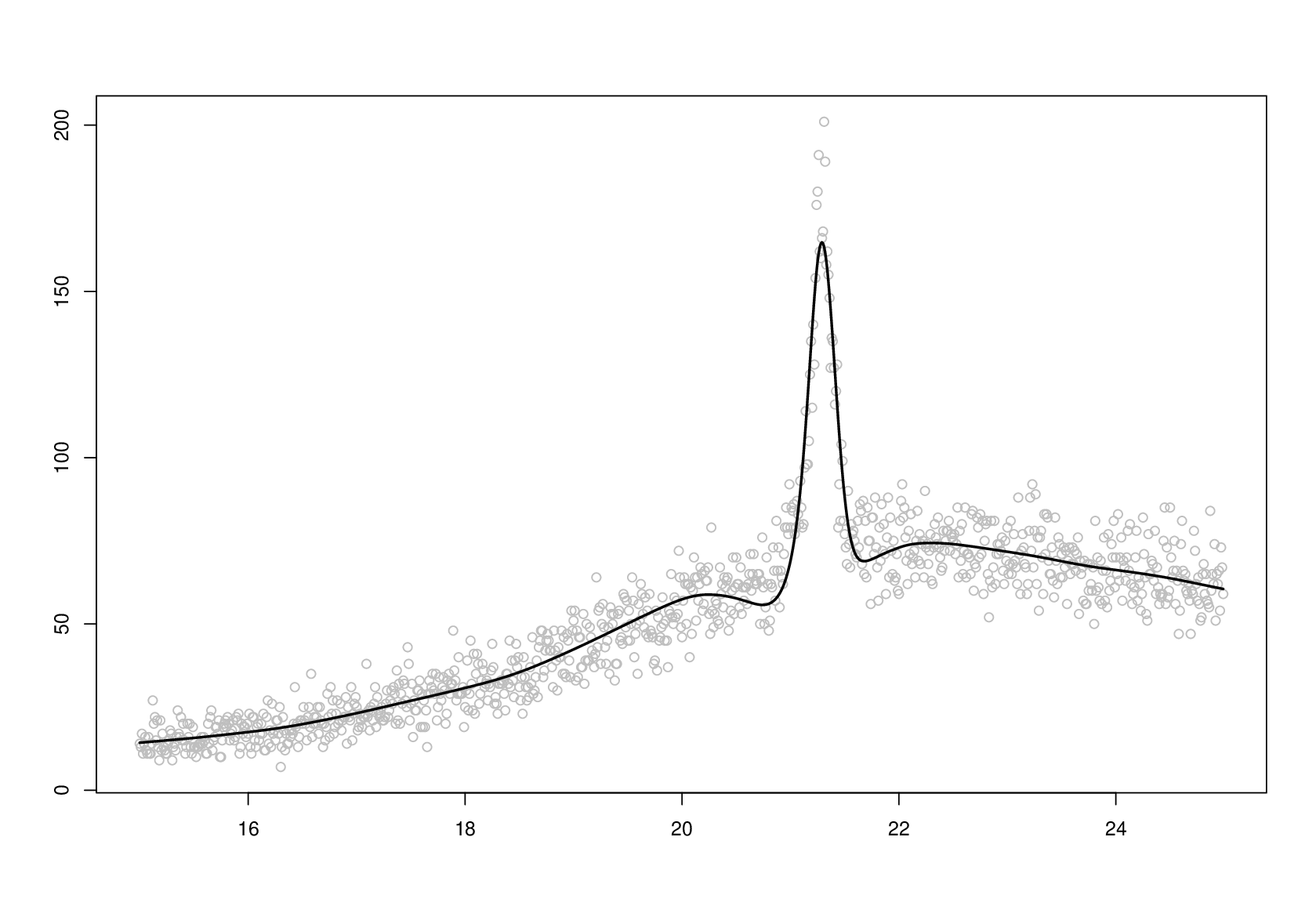,angle=270,width=7cm}
  \psfig{file=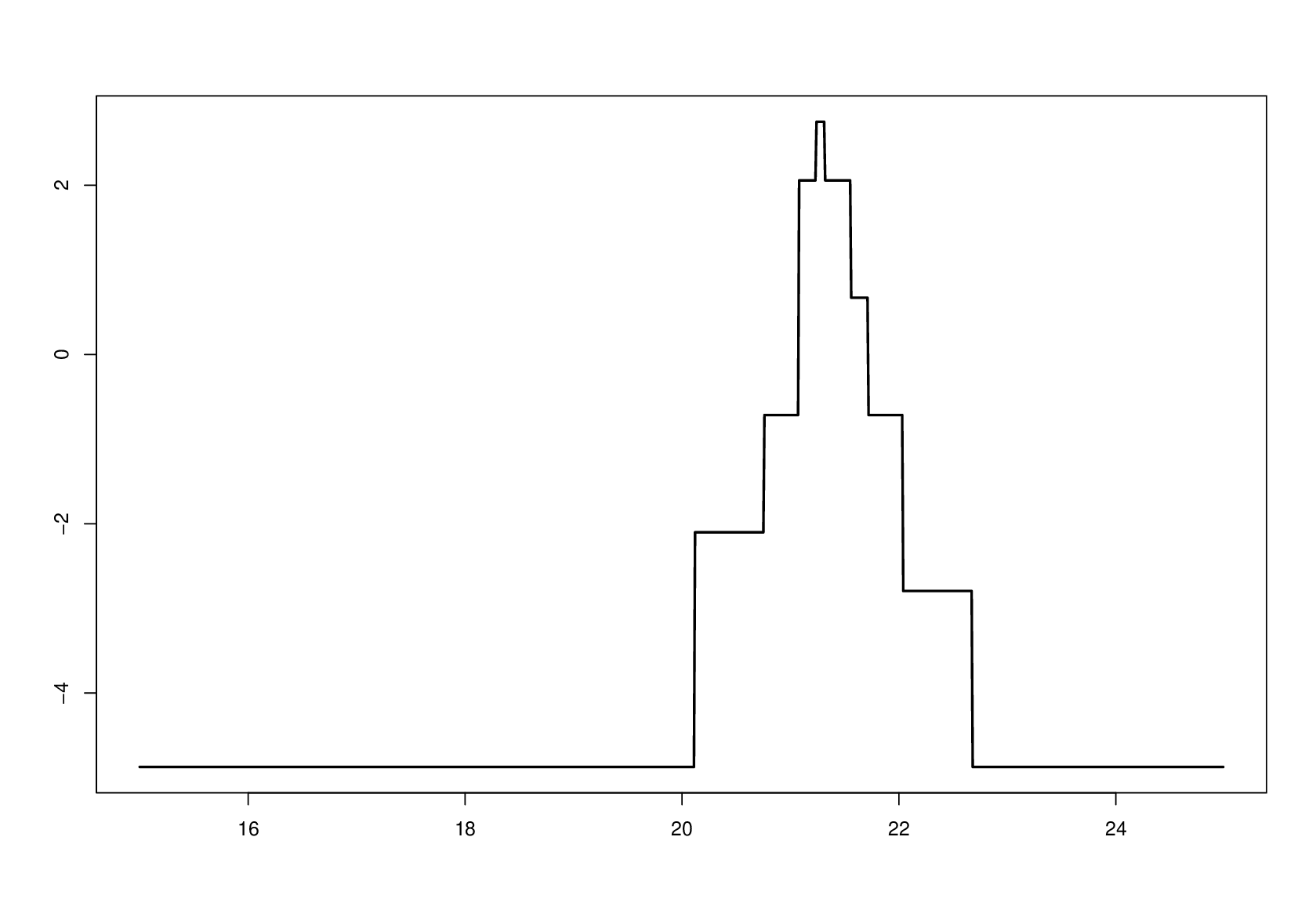,angle=270,width=7cm}\\
  \caption{{\footnotesize The rows show the results after 1, 15, 25
      and 33 iterations of the weighted smoothing splines procedure to
      the first 1000 data points of the thin film data. The left panel
      shows the curve and the right panel the weights $\lambda_i$ on a
      logarithmic scale.} \label{wssiters}} 
\end{figure}

We apply the procedure to the thin film data of Figure
\ref{FIGm35020a}. The value of $\sigma_n$ of (\ref{sigma1}) is 8.3868. With
$n=7001$ and $\tau_n(\alpha)=3$ we have 
\[\sigma_n\sqrt{\tau_n(\alpha)\log n}=8.3868\sqrt{3\cdot \log
  7001}=43.22\]
The upper panel of Figure \ref{wssthinfilmw} shows the resulting
curve: the lower panel shows the associated values of the $\lambda_i$ on a
logarithmic scale. It is noticeable that the values of the $\lambda_i$
are large in the neighbourhoods of the large peaks and small outside
of these. The manner in which the curve alters in the course of the
iterations is shown on a larger scale in Figure \ref{wssiters}. The
rows show the results after 1, 15 and 25 iterations and the final
result after 33 iterations for the first 1000 observations. In each
case the left panel shows the curve and the right panel the weights
$\lambda_i$ on a logarithmic scale. Initially the weights
are constant with a value of $2.9\cdot 10^{-8}.$ After 15 iterations
they are still constant but now with the common value $2.5\cdot
10^{-4}.$ After 25 iterations the smallest weights are $7.4\cdot
10^{-3}$ and the largest is 0.18. The smallest weights for the final
curve are still $7.4\cdot 10^{-3}$ but the largest weight is now
20. The values of the $\lambda_i$ differ by a factor of 30000. The
final row shows the advantage of the local weights $\lambda_i.$ 
Where the data can be fitted with a smooth curve the $\lambda_i$ are
small and the fit is smooth. Where there is  a pronounced peak the
values of the $\lambda_i$ are large and this forces the solution of
(\ref{eq:SS}) to adjust to the peak. 

\section{Examples and simulations  \label{examples}}
\subsection{The thin film data}
\begin{figure}
  \centering
  \psfig{file=FIG_XRAY_wss.ps,angle=270,width=7cm}
  \psfig{file=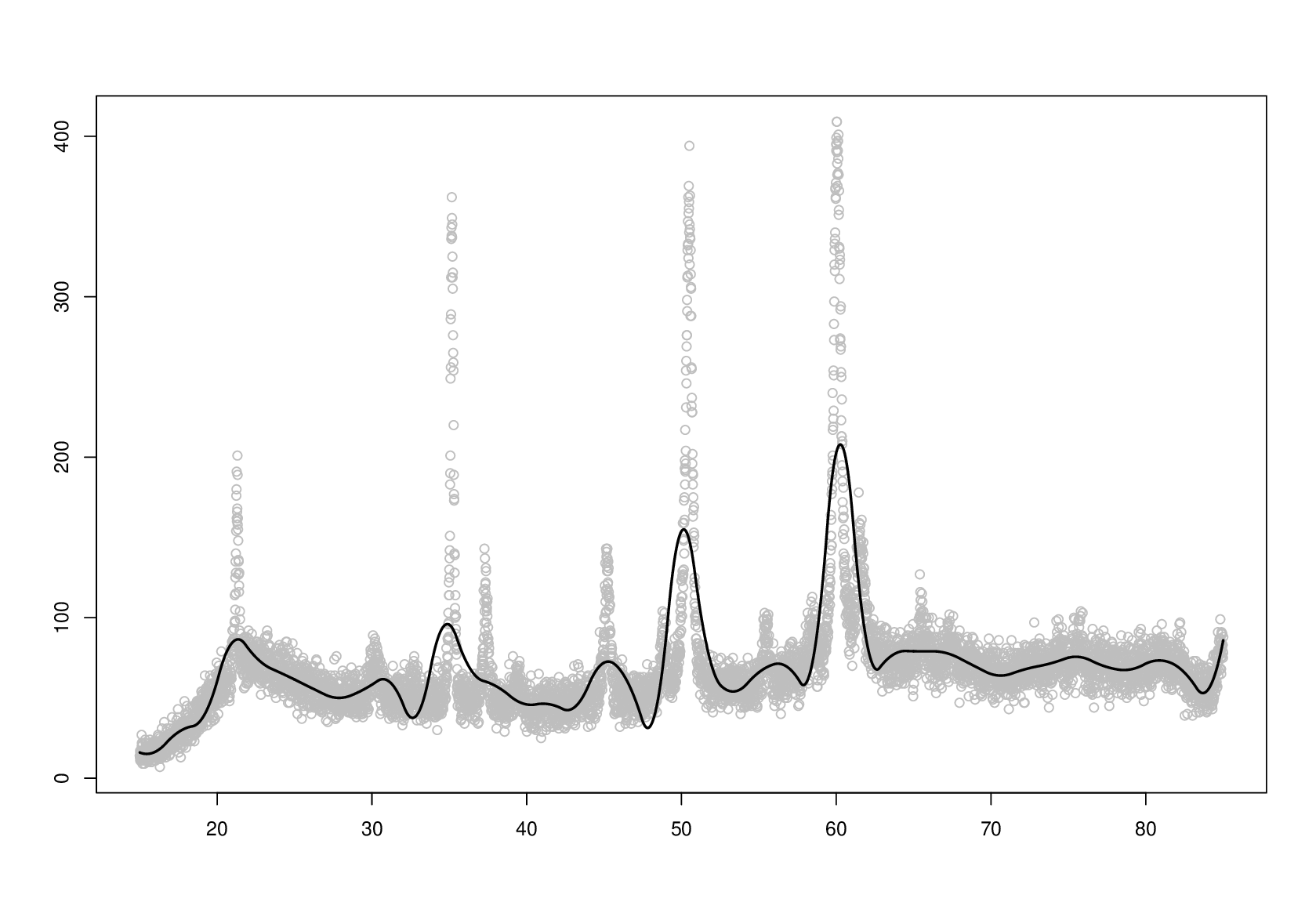,angle=270,width=7cm}\\
  \psfig{file=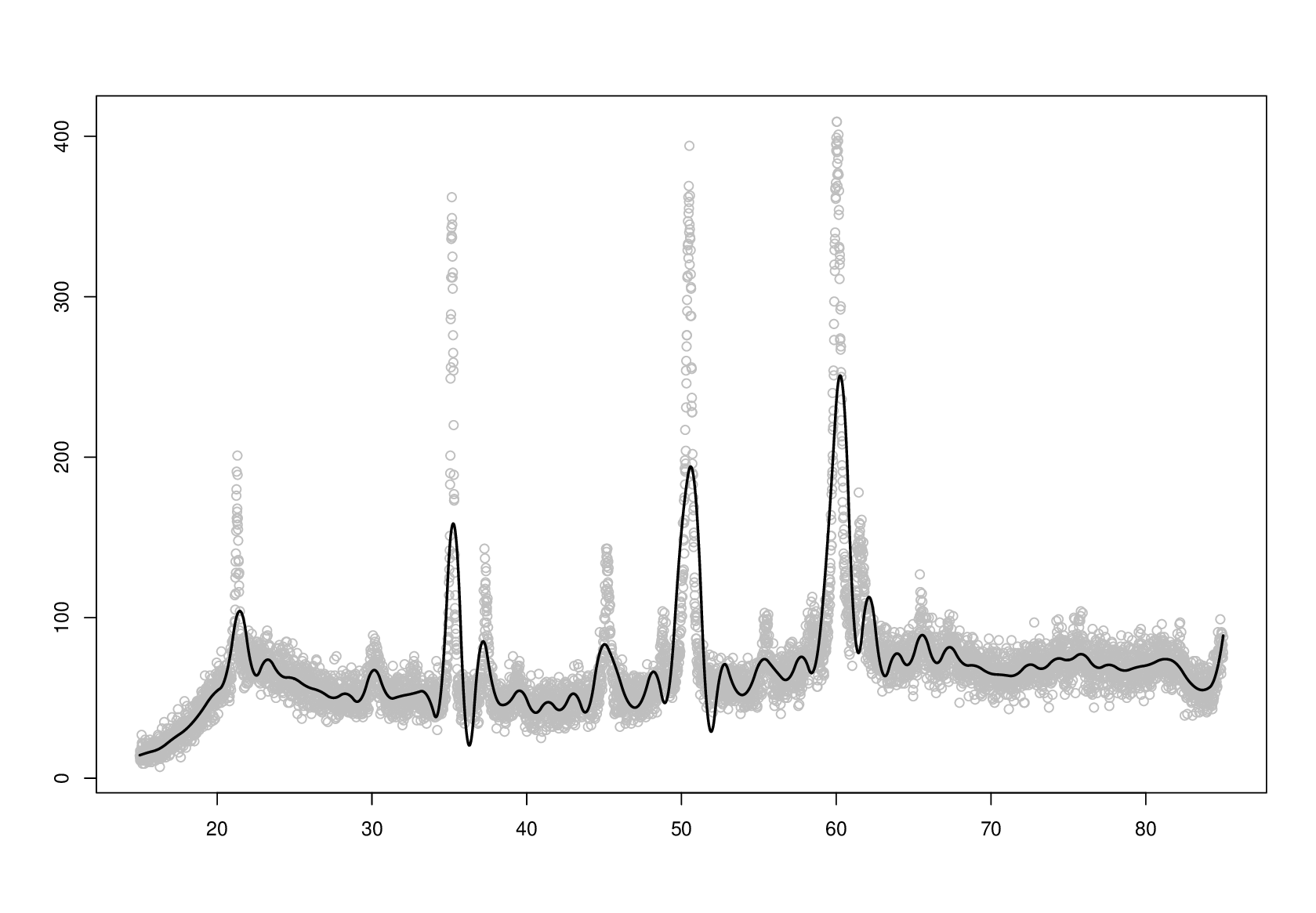,angle=270,width=7cm}
  \psfig{file=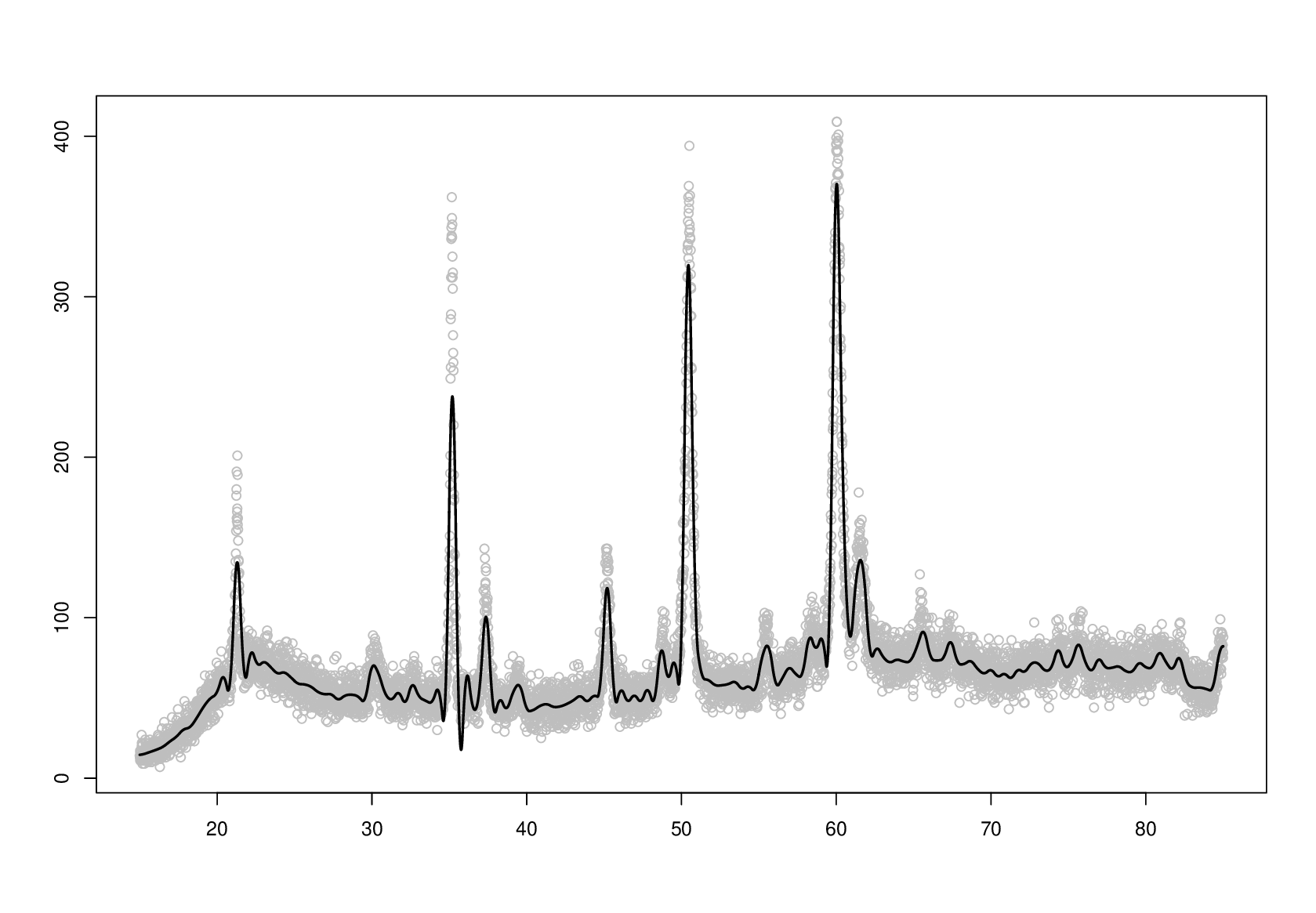,angle=270,width=7cm}\\
  \psfig{file=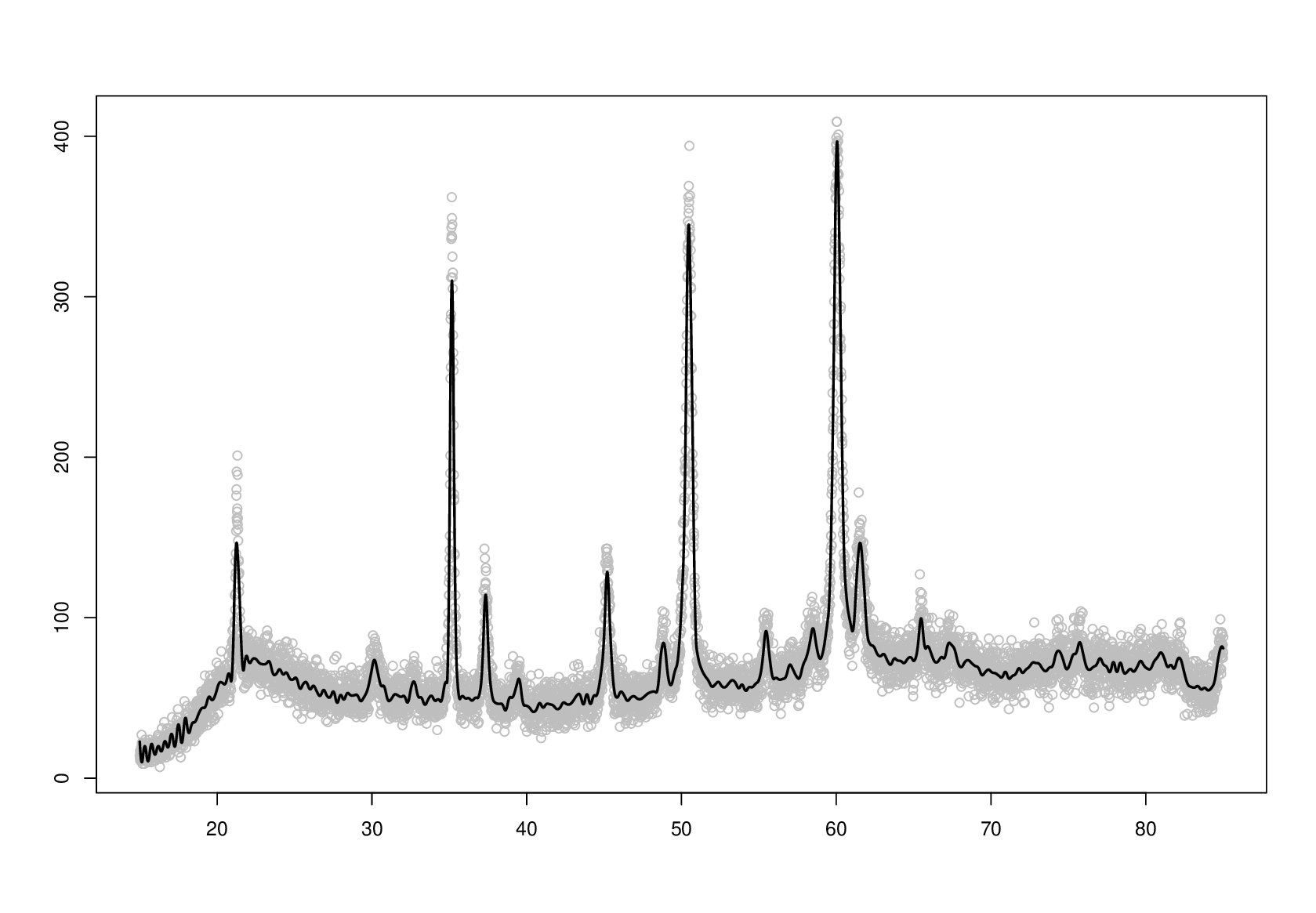,angle=270,width=7cm}
  \psfig{file=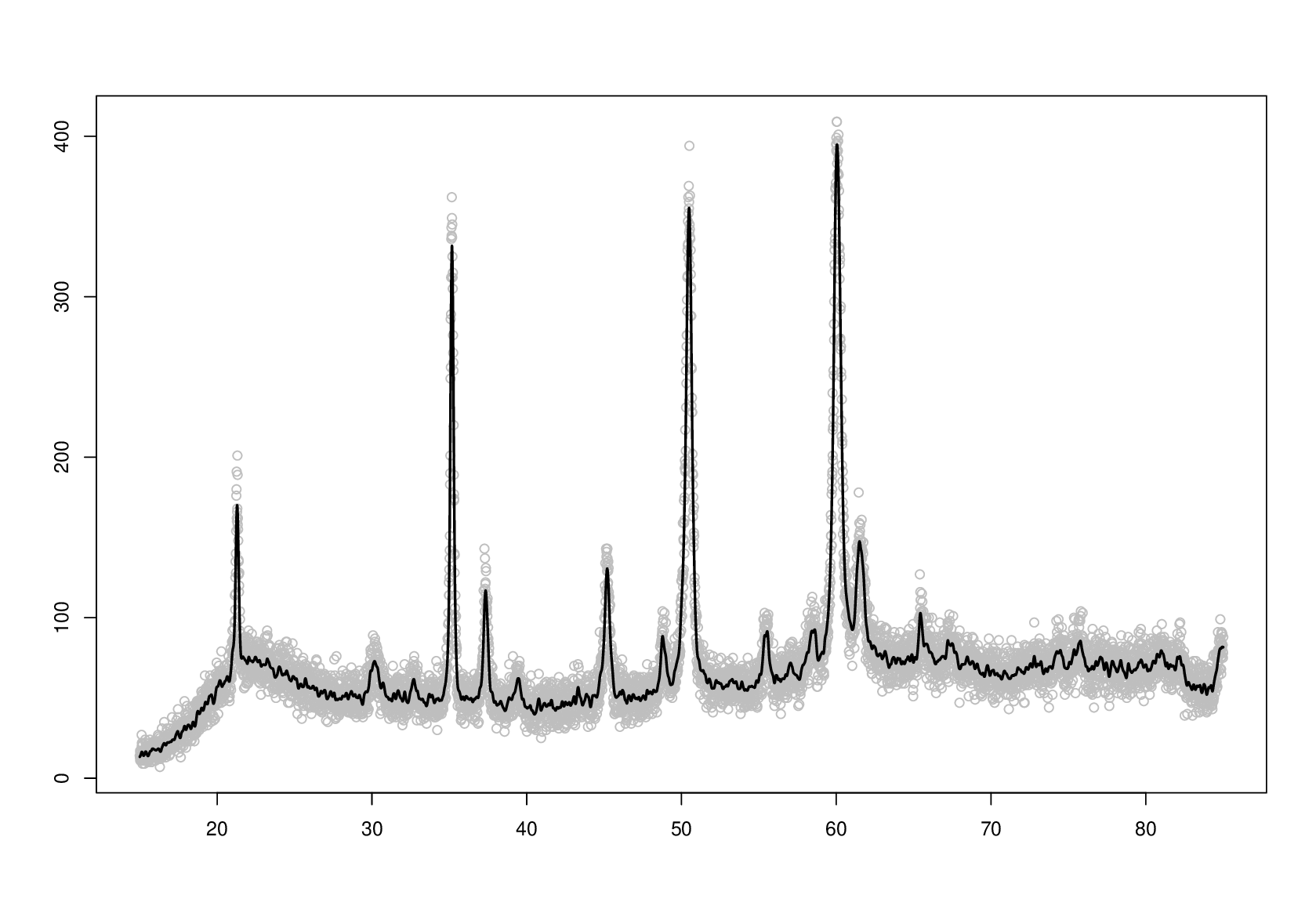,angle=270,width=7cm}\\
  \caption{{\footnotesize The top row shows from left to right the
      {\it wss} and {\it pspl40} reconstructions, the centre row 
      the results for the  {\it pspl80} and  {\it pspl160}
      reconstructions and the bottom row the results for the {\it
        pspl320} and the {\it smspl}
      reconstructions}. \label{FIGm35020asmth}} 
\end{figure}

\begin{figure}
  \centering
  \psfig{file=FIG_XRAY_wss_part.ps,height=.7\textwidth,width=.3\textwidth,angle=270}
  \psfig{file=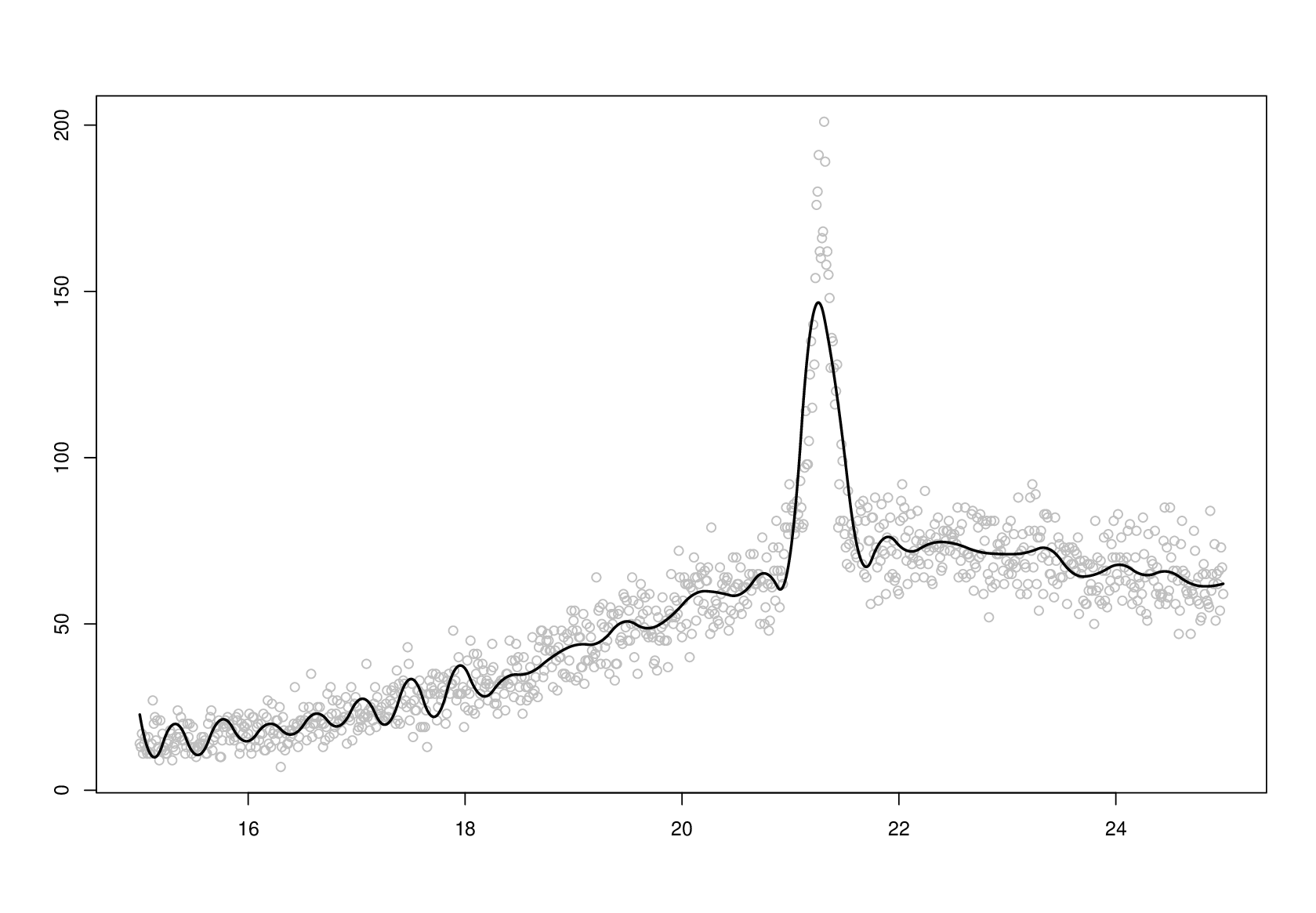,height=.7\textwidth,width=.3\textwidth,angle=270}
  \psfig{file=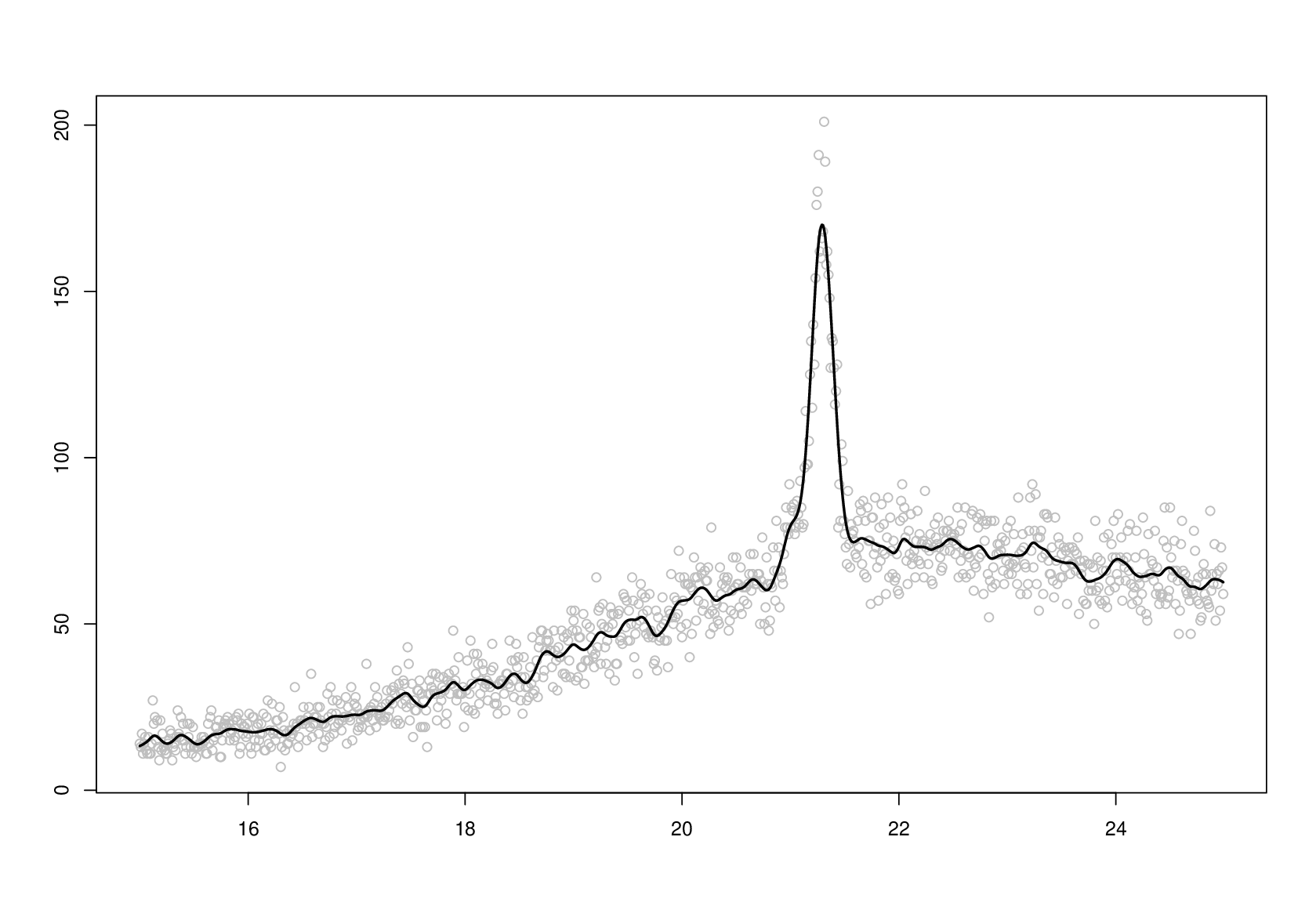,height=.7\textwidth,width=.3\textwidth,angle=270}\\

  \caption{ {\footnotesize The first 1000 observation of the thin film
      data with the (from top to bottom) {\it wss, pspl320} and {\it smspl}
    reconstructions.} \label{FIGm35020asmthpart}}
\end{figure}

 The estimators we consider are the weighted 
smoothing spline ({\it wss}),  the spatially adaptive spline method
due to \cite{RUPPCARR00} and the standard
smoothing spline ({\it smspl}) with smoothing parameter chosen by
cross validation. The Ruppert--Carroll method uses so called
`penalized splines' which are the p-splines of \cite{EILMAR96} (see
also \cite{OSUL86,OSUL88}). In contrast to smoothing splines they use
a spatially weighted penalty term with the weights being determined by
generalized cross-validation. The method is not fully automatic
and requires the specification of the maximum number of knots. Based
on  \cite{RUPPCARR00} the numbers we choose are 40, 80, 160 and 320: we 
denote the corresponding estimators by {\it pspl40, pspl80, pspl160}
and {\it   pspl320}. Figure \ref{FIGm35020asmth} shows the
results for the complete data set. It is seen that the peaks are
satisfactorily captured only by the {\it wss, pspl320 and smspl}
reconstructions. Figure \ref{FIGm35020asmthpart} shows the results for
the first 1000 observations only for these three methods. Only the
{\it wss} succeeds in capturing the peaks and giving a smooth reconstruction
between the peaks.

\subsection{Some simulation results}
We give the results of a small simulation study using the functions of
\cite{RUPPCARR00}
\begin{equation*} \label{rupcarfn}
f(x)=f(x;j)=\sqrt{x(1-x)}
\sin\left(\frac{2\pi(1+2^{(9-4j)/5)}}{x+2^{(9-4j)/5}}
\right)
\end{equation*} 
with $j=6$ and the bumps data of \cite{DONJOH95}. We
consider signal to noise ratios of 3 and 7. The tables 
gives the median (MRISE) of the root integrated square error
\[ \text{RISE}(f, {\hat f}_n)=\left(\int_0^1 (f(t)-{\hat f}_n(t))^2\,
  dt\right)^{1/2}\]for
the fit itself and the first and second derivatives
$\text{MRISE}(f^{(i)},{\hat f}_n^{(i)}), i=1,2.$ The results are 
based on 1000 simulations.

\begin{table}
\begin{center}
\begin{tabular}{|l|llll|llll|llll|}
\hline
\multicolumn{13}{|c|}{ $\sigma=0.288/7\approx 0.0411$}\\
\hline
&\multicolumn{4}{c|}{Fit}&\multicolumn{4}{c|}{First
  derivative}&\multicolumn{4}{c|}{Second derivative}\\
&400&800&1600&3200&400&800&1600&3200&400&800&1600&3200\\
\hline
wss&0.030&0.021&0.016&0.012&0.267&0.161&0.096&0.057&3.98&1.88&0.85&0.37\\ 
pspl40&0.062&0.058&0.057&0.056&0.559&0.388&0.274&0.192&6.48&3.33&1.69&0.85\\ 
pspl80&0.027&0.021&0.017&0.015&0.339&0.218&0.145&0.101&5.26&2.69&1.32&0.67\\ 
pspl160&0.022&0.016&0.012&0.009&0.244&0.148&0.090&0.054&3.87&2.12&1.14&0.56\\ 
smspl&0.024&0.016&0.012&0.009&0.294&0.139&0.080&0.047&4.77&1.77&0.79&0.37\\ 
\hline
\multicolumn{13}{|c|}{$\sigma=0.288/3\approx 0.096$}\\
\hline
&\multicolumn{4}{c|}{Fit}&\multicolumn{4}{c|}{First
  derivative}&\multicolumn{4}{c|}{Second derivative}\\
&400&800&1600&3200&400&800&1600&3200&400&800&1600&3200\\
\hline
wss&5.60&2.71&1.22&0.54&0.417&0.244&0.150&0.093&5.60&2.71&1.22&0.54\\ 
pspl40&6.49&3.34&1.69&0.85&0.567&0.392&0.275&0.193&6.49&3.34&1.69&0.85\\ 
pspl80&5.59&2.79&1.35&0.68&0.414&0.256&0.159&0.106&5.59&2.79&1.35&0.68\\ 
pspl160&5.19&2.59&1.25&0.62&0.387&0.242&0.142&0.083&5.19&2.59&1.25&0.62\\ 
smspl&5.48&2.43&1.11&0.52&0.401&0.232&0.136&0.080&5.48&2.43&1.11&0.52\\ 
\hline
\end{tabular}
\caption{Values of the MRISE based on 1000 simulations
    for the Ruppert and Carroll function with
    $j=6$.\label{rupcarimse}} 
\end{center}
\end{table}

\begin{table}
\begin{center}

\begin{tabular}{|l|llll|llll|llll|}
\hline
\multicolumn{13}{|c|}{$\sigma= 2.2/3 \approx 0.733$}\\
\hline
&\multicolumn{4}{c|}{Fit}&\multicolumn{4}{c|}{First
  derivative}&\multicolumn{4}{c|}{Second derivative}\\
&400&800&1600&3200&400&800&1600&3200&400&800&1600&3200\\
\hline
wss&0.80&0.69&0.56&0.43&18.9&18.4&14.7&10.6&637&788&761&642\\ 
pspl 40&1.55&1.54&1.52&1.52&31.1&27.6&21.9&16.5&900&959&860&693\\ 
pspl 80&1.18&1.15&1.14&1.14&29.1&26.4&21.2&16.0&889&957&860&693\\ 
pspl 160&0.84&0.81&0.79&0.78&24.1&23.9&19.8&15.1&811&942&857&692\\ 
smspl&1.14&0.91&0.84&0.65&28.8&24.9&20.1&14.4&890&951&858&690\\ 
\hline
\multicolumn{13}{|c|}{$\sigma= 2.2/7 \approx 0.314$}\\
\hline
&\multicolumn{4}{c|}{Fit}&\multicolumn{4}{c|}{First
  derivative}&\multicolumn{4}{c|}{Second derivative}\\
&400&800&1600&3200&400&800&1600&3200&400&800&1600&3200\\
\hline
wss&0.44&0.35&0.25&0.18&11.9&11.2&9.3&7.2&417&522&570&539 \\ 
pspl 40&1.54&1.53&1.52&1.52&31.1&27.6&21.9&16.5&900&959&860&693\\ 
pspl 80&1.14&1.13&1.13&1.13&29.0&26.3&21.2&16.0&889&957&860&692\\ 
pspl 160&0.74&0.76&0.76&0.77&23.4&23.7&19.7&15.1&801&941&857&692\\ 
smspl&1.10&0.86&0.81&0.62&28.7&24.8&20.1&14.3&889&950&858&690\\ 
\hline
\end{tabular}
\caption{Values of the MRISE based on 1000 simulations
    for the bumps 
  function of \cite{DONJOH95}. \label{bumpsimse}} 

\end{center}
\end{table}
 We expect locally adaptive methods to perform better when the signal exhibits
large changes in local variability and the signal to noise
ratio is large. This is borne out by the results. The local
variability of the Ruppert-Carroll is not large and there is not much
to choose between the four methods {\it  wss, pspl80,pspl160} and {\it
  smspl} both in the low and high signal to noise scenarios. However
the {RMISE often disguises clear differences in the behaviour of the
  estimators. Figure  \ref{FIG_RupCar} shows a typical result for the
  high signal to   noise regime for the  Ruppert-Carroll-function and
  a sample size $n=1600.$ 
\begin{figure}
  \centering
\psfig{file=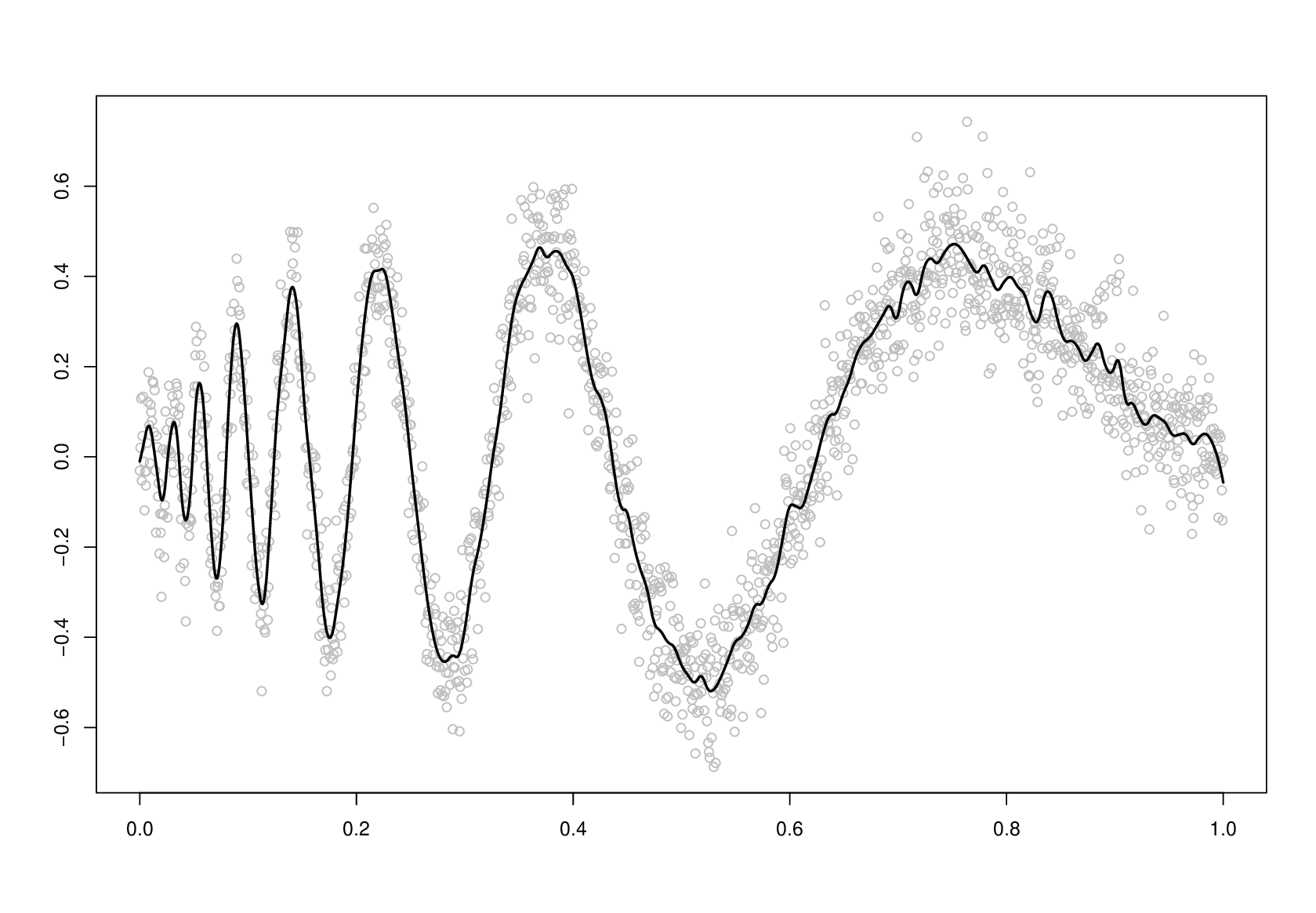,height=.7\textwidth,
  width=.3\textwidth,angle=270} 
  \psfig{file=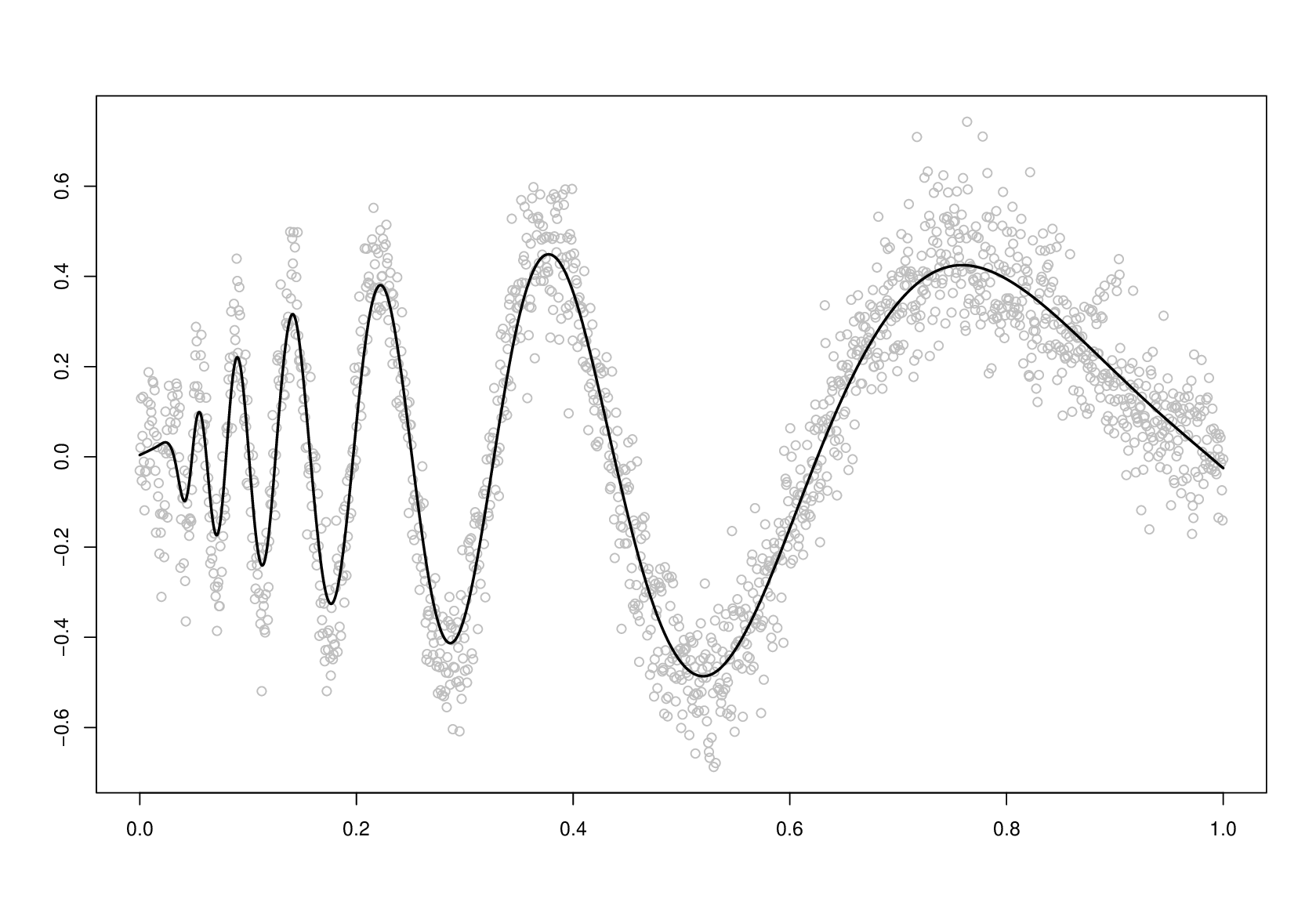,height=.7\textwidth,
    width=.3\textwidth,angle=270} 
  \caption{{\footnotesize The top panel shows a reconstruction using
      {\it pspl160} and the bottom panel shows the {\it wss}
      reconstruction for the same data. The sample size is $n=1600.$}
      \label{FIG_RupCar}} 
\end{figure}

The local variability of the bumps data is much more pronounced and
the {\it wss} estimator outperforms the other estimators in all cases.

\section{Heteroscedasticity and robustness}\label{sclhetsced}
\subsection{Nonparametric scale approximations} \label{nonparvol}
The ideas developed in the previous section can also be used to obtain
nonparametric approximations to heteroscedastic noise. The model we
use is
\begin{eqnarray} \label{scalemod}
Y(t)=\sigma(t)Z(t),\quad 0 \le t \le 1,
\end{eqnarray}
$Z(t)$ Gaussian white noise. Given data $(t_i, Y(t_i))$, $i=1, \dots,
n$, we define a confidence region as follows. We define for a function
$s:[0,\,1] \rightarrow (0,\infty)$ and an interval $I \subset [0,\,1]$ 
\begin{eqnarray*} \label{defvol}
  v({\mathbold Y}_n,I,s)=\sum_{t_i\in I}Y(t_i)^2/s(t_i)^2
\end{eqnarray*}
and then set
\begin{eqnarray*}
{\mathcal C}_n({\mathbold Y}_n,{\mathcal I}_n,\alpha_n)
&=&\{s:\text{qu}((1-\alpha_n)/2,\vert I\vert) \le v({\mathbold
  Y}_n,I,s)\label{confregsig}\\ 
&& \quad\quad\quad\le
\text{qu}((1+\alpha_n)/2,\vert I\vert) ,\quad I \in {\mathcal I}_n\}\nonumber
\end{eqnarray*}
where $\text{qu}(\gamma,k)$ denotes the $\gamma$--quantile of the 
chi--squared distribution with $k$ degrees of freedom. The rationale
is clear. Under the model (\ref{scalemod}) the $v({\mathbold
  Y}_n,I,\sigma)$ has the chi--squared distribution
with $\vert I\vert$ degrees of freedom. By an appropriate choice of
$\alpha_n,$ which may be determined by simulations, ${\mathcal
  C}_n({\mathbold Y}_n,{\mathcal I}_n,\alpha_n)$ is an
$\alpha$--confidence region for $\sigma$:
\begin{eqnarray*}
P( \sigma \in  {\mathcal
  C}_n({\mathbold Y}_n,{\mathcal I}_n,\alpha_n))=\alpha
\end{eqnarray*}
so that the confidence region is uniform, exact and
non-asymptotic. Furthermore in this particular model there are no
``nuisance'' parameters corresponding to the $\sigma$ of model
(\ref{basicmod}). The default value of $\gamma_n$ we use is
\begin{eqnarray*} \label{gammdef}
\gamma_n = 1-\exp(-1.5\log(n))=1-n^{-1.5}
\end{eqnarray*}
which roughly corresponds to the default choice of $\tau_n=3$ in the
definition of ${\mathcal A}_n.$ As before the second step is to
regularize in ${\mathcal C}_n({\mathbold Y}_n,{\mathcal
  I}_n,\alpha_n).$ One possibility which is useful for quantifying the
changes in volatility of financial data, the volatility of the
volatility, is to take $s$ to be piecewise constant and to minimize
the number of intervals of constancy (see \cite{DAV06}). In the
present context however we are looking for a smooth approximation and we 
take recourse to weighted smoothing  
splines. We take $s=s_n$ to be the solution of 
\begin{eqnarray*} \label{sclssp}
\min \quad \sum_{i=1}^n \lambda_i(\vert y_i\vert-s_n(t_i))^2 +
\int_0^1 s_n^{(2)}(t)^2\,dt. 
\end{eqnarray*}
where again the local weights are data dependent and are chosen so that the
solution $s_n$ lies in ${\mathcal C}_n({\mathbold Y}_n,{\mathcal
  I}_n,\alpha_n).$  The procedure we use is similar to that described
in Section \ref{wssproc} but with some modifications. On intervals $I$
where the inequality 
\begin{eqnarray} \label{approxvol1}
\text{qu}((1-\alpha_n)/2,\vert I\vert) \le v({\mathbold
  y}_n,I,s_n) \le \text{qu}((1+\alpha_n)/2,\vert I\vert)
\end{eqnarray}
is not satisfied we increase the weights by a factor of $q$ but we do
this firstly for single observations, that is intervals of length
one. When (\ref{approxvol1}) is satisfied for all such intervals we
consider intervals of length two. When again all the inequalities are
satisfied we move on to the next longer intervals until finally all
inequalities are satisfied. A similar procedure was used in
\cite{DAVKOV04} in the context of approximating spectral 
densities. Figure \ref{fig:vardata} shows the result of the procedure
applied to data generated according to the model
\begin{eqnarray*} \label{sinevol}
  Y(t)=\sin(4\pi t)^2Z(t).
\end{eqnarray*}
\begin{figure}
  \centering
  \psfig{file=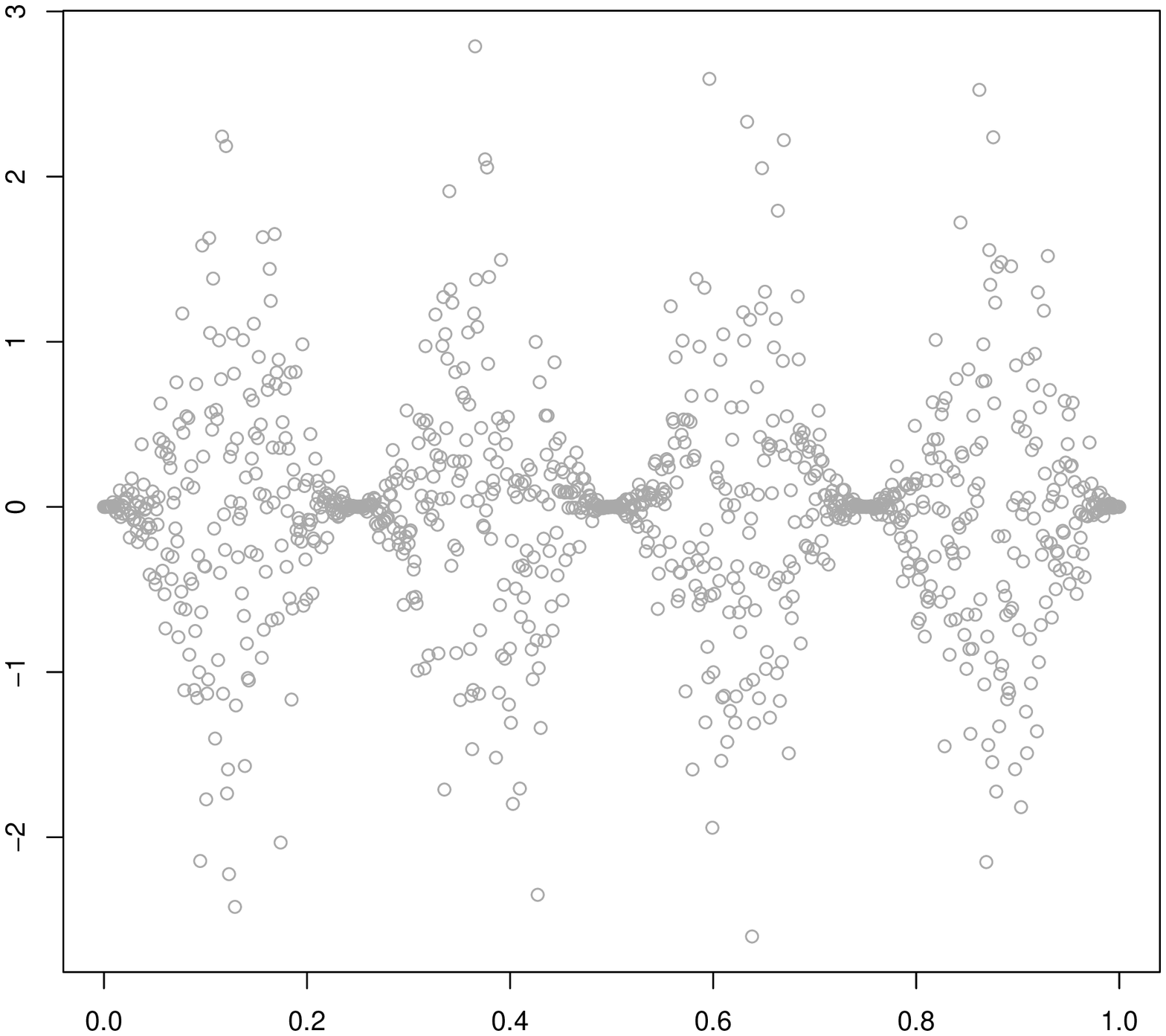,height=.7\textwidth,width=.3\textwidth,angle=270}
  \psfig{file=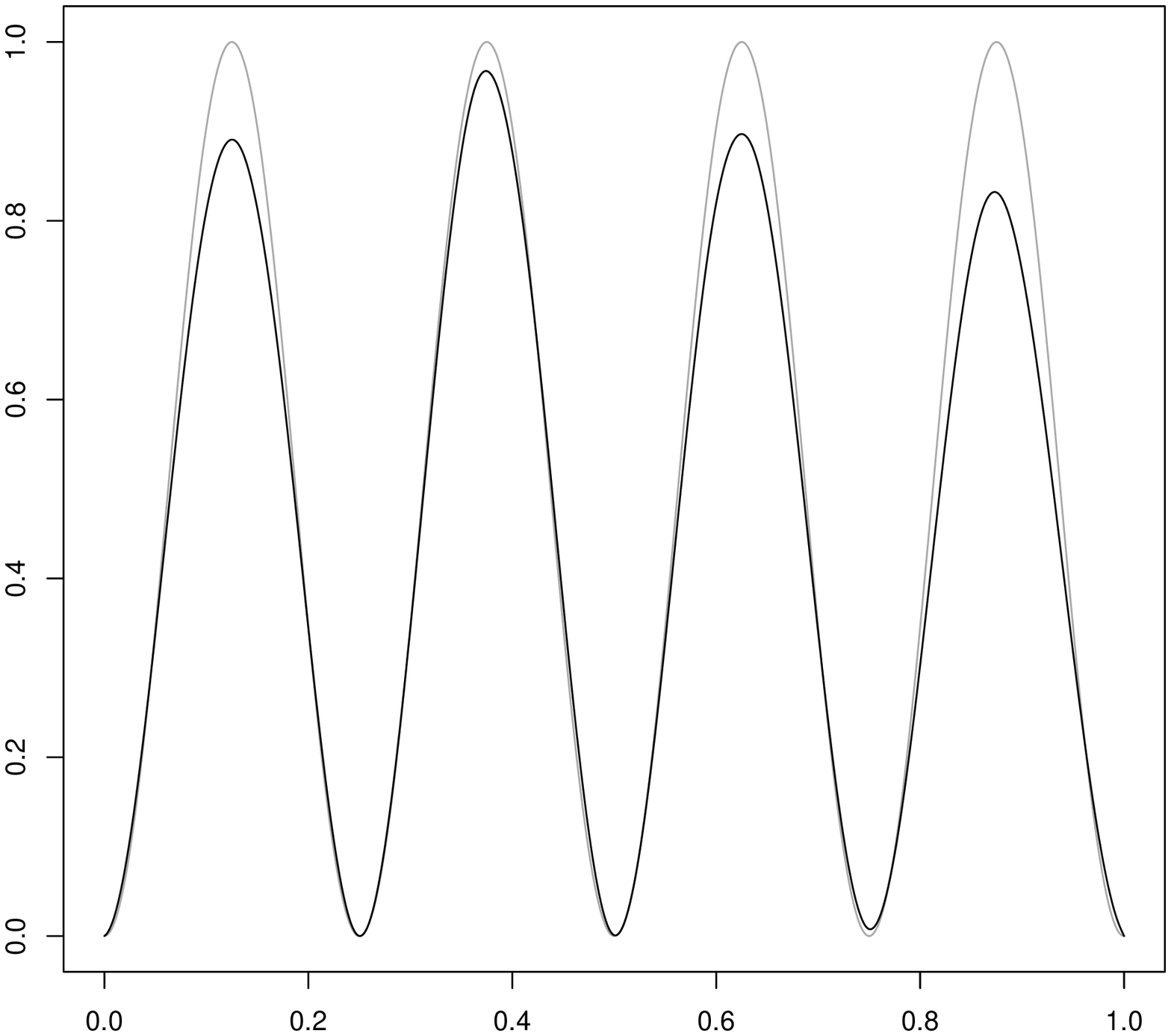,height=.7\textwidth,width=.3\textwidth,angle=270}
   \caption{{\footnotesize Top panel: heteroscedastic noise. Bottom
       panel: the scale function and its reconstruction using weighted
       smoothing spline.}. \label{fig:vardata}} 
\end{figure}

\subsection{Robust smoothing}
A complete robustification of the procedure described in Section
\ref{wssproc} would entail replacing (\ref{eq:WSS}) by, for example,
\begin{eqnarray*} \label{eq:WSSrob}
 \min S(g,{\mathbold\lambda}):=
\sum_{i=1}^n\lambda_i\vert y(t_i)-g(t_i)\vert+\int_0^1g^{(2)}(t)^2\,dt,   
\end{eqnarray*}
and the definition of approximation (\ref{defwifn}) by 
\begin{eqnarray*} \label{defwifnrob}
{\tilde w}({\mathbold y}_n,I,g)=\frac{1}{\sqrt{\vert I \vert}}\,\sum_{t_i\in
  I}\text{sgn}(r({\mathbold y}_n,t_i,g))
\end{eqnarray*}
to give rise to the confidence region 
\begin{eqnarray*} \label{approxreg3}
{\mathcal D}_n({\mathbold y}_n, {\mathcal I}_n) = \big\{g:  \max_{ I
  \in {\mathcal I}_n} \,{\tilde w}({\mathbold y}_n,I,g) \le
\sqrt{2\log n\,}\,\big\}
\end{eqnarray*}
(see \cite{Duemkov05}). A much simpler but reasonably
effective method is the following. The noise level $\sigma_n$ is
quantified by (\ref{sigma1}). A running median with a window width of
say five observations is applied to the data 
\[m_5(t_i):= \text{MED}(y(t_{i-2}),y(t_{i-1}),y(t_i),y(t_{i+1}),y(t_{i+2}))\]
and any data point $y(t_i)$ for which 
\[\vert y(t_i)-m_5(t_i)\vert \ge 3.5\sigma_n,\]
is replaced by $m_5(t_i)$ (see \cite{HAM85}). The
weighted splines procedure is now applied to the cleaned data set. The
procedure will work well as long as no group of five successive
observations contains more than two outliers. Figure
\ref{fig:robwssWSS} shows the result of applying this robustified
procedure to a sine curve contaminated with Cauchy noise.
 
\begin{figure}
  \centering
  \psfig{file=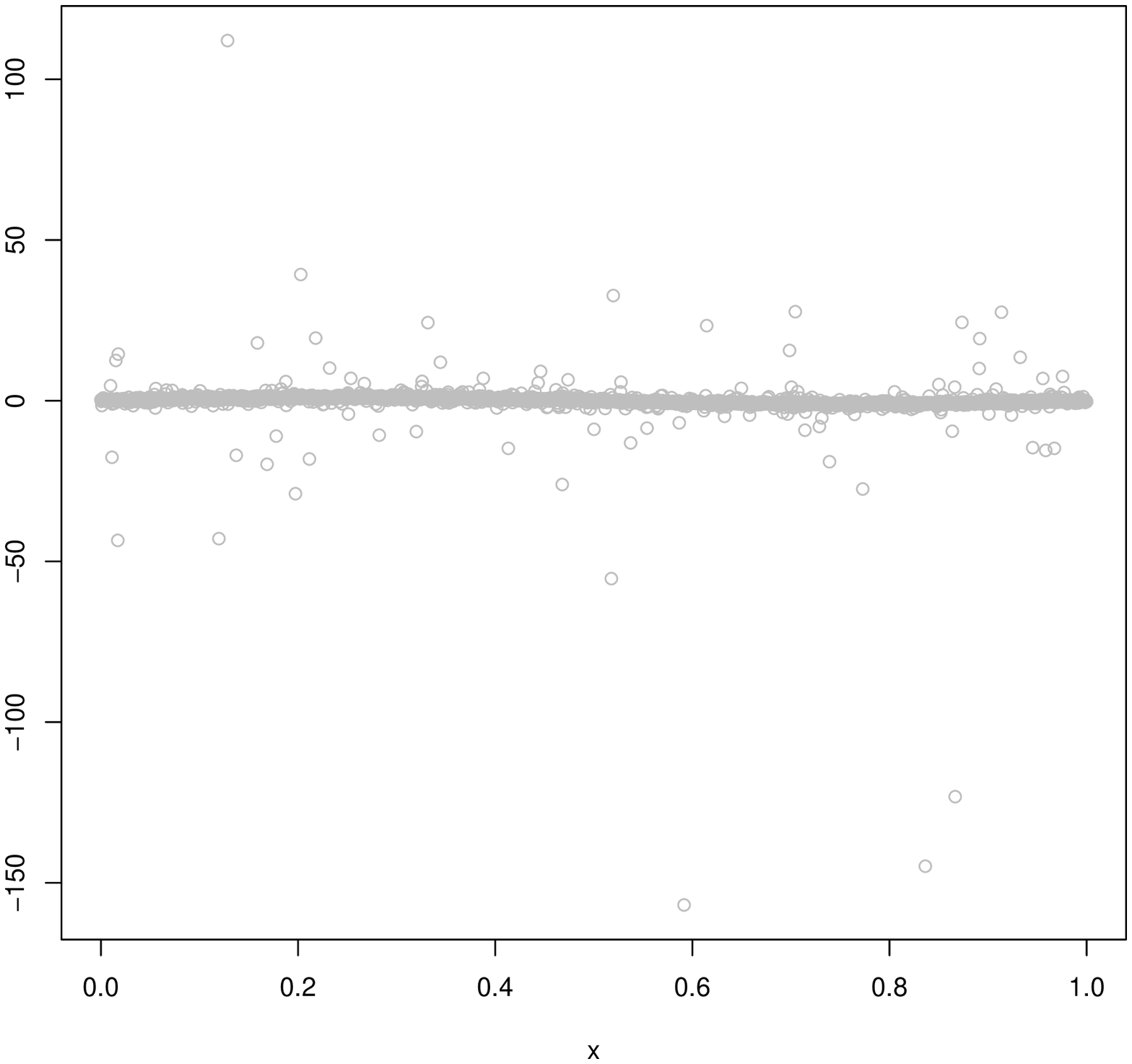,height=.7\textwidth,width=.3\textwidth,angle=270}
  \psfig{file=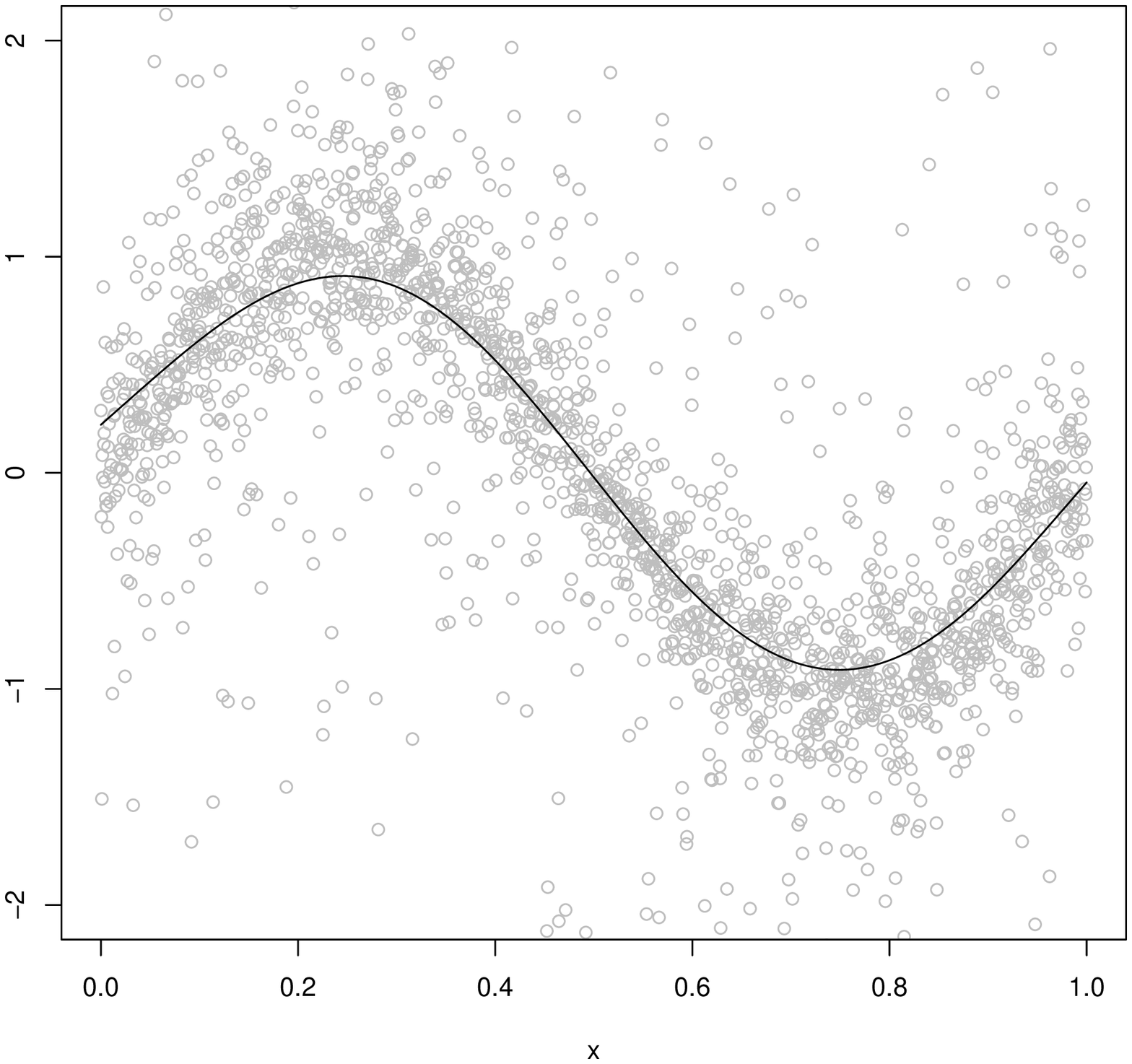,height=.7\textwidth,width=.3\textwidth,angle=270}
  
  \caption{{\footnotesize The robustified weighted spline procedure
      applied to a sine curve contaminated with Cauchy noise}}
  \label{fig:robwssWSS}
\end{figure}

\section{Image analysis and weighted thin plate smoothing splines}
\label{imageanal} 
\subsection{Weighted thin plate smoothing splines}
We consider data ${\mathbold y}_n=\{({\mathbold t}_i,y({\mathbold
  t}_i)):\,i=1,\ldots, n^2\,\} $  with the ${\mathbold t}_i$ of the form 
\[{\mathbold t}_i=(j_i/n,k_i/n), \quad j_i,k_i =0,\ldots,n-1.\]
Corresponding to (\ref{eq:WSS}) we consider minimizing
 \begin{eqnarray*} \label{eq:WTPS}
S(f,{\mathbold\lambda}):=\sum_{i=1}^{n^2}\lambda({\mathbold t}_i)(y({\mathbold
  t_i})-f({\mathbold t_i}))^2+J(f)
\end{eqnarray*}
with
\begin{eqnarray*}
  J(f)=\int_0^1\int_0^1\left(\left(\frac{\partial^2 f}{\partial^2
        s}\right)^2+2\left(\frac{\partial^2 f}{\partial s\partial
        t}\right)^2+\left(\frac{\partial^2 f}{\partial^2
        t}\right)^2\right)dsdt  
\end{eqnarray*}
It can be shown that the solution is a natural thin plate spline. We
refer to \cite{GRESILV94}.

\subsection{Approximation in two dimensions}
For a given function $g:[0,\,1]^2\rightarrow \rz$ and a family
${\mathcal G}_n$ of subsets $G$ of $[0,\,1]^2$ we define 
\begin{eqnarray*} \label{defwifnim}
w({\mathbold y}_n,G,g)=\frac{1}{\sqrt{\vert G \vert}}\,\sum_{{\mathbold t}_i\in
  G}(y_n({\mathbold t}_i)-g({\mathbold t}_i)).
\end{eqnarray*}
For data generated by the model
\begin{eqnarray*} \label{immod}
Y({\mathbold t})=f({\mathbold t})+\sigma Z({\mathbold t}),\quad
{\mathbold t} \in [0,\,1]^2
\end{eqnarray*}
this leads to the confidence region
\begin{eqnarray*} \label{confregim}
\lefteqn{{\mathcal H}^*({\mathbold Y}_n,{\mathcal G}_n,\tau)}\\
&=&\{g: \max_{ G\in
  {\mathcal G}_n}\,\vert w({\mathbold Y}_n,G,g)\vert \le \sigma_n
\sqrt{2 \tau \log(n)}\,\}. \nonumber
\end{eqnarray*}
The additional factor 2 is due to the fact that we now have $n^2$
observations. The noise level $\sigma_n$ is defined  by
\begin{eqnarray*}
\sigma_n&=&\frac{1.48}{2}\text{MED}\Big(\big\{\big|y(\textstyle
\frac{j_i+1}{n},\frac{k_i+1}{n})-y(\frac{j_i+1}{n},\frac{k_i}{n})\\
&&-y(\frac{j_i}{n},
\frac{k_i+1}{n})+y(\frac{j_i}{n},\frac{k_i}{n})\big|:i=1,\dots, 
n^2\big\}\Big).\nonumber  
 \label{noiseim}
\end{eqnarray*}
The quality of the results depends on the choice of ${\mathcal G}_n$. If
${\mathcal G}_n$ contains too few sets then the concept of approximation
is too crude. Consequently we require a fine division of $[0,\,1]^2$
but one which allows the $w({\mathbold y}_n,G,g)$ 
to be efficiently calculated. Work in this direction has been done
and we refer to \cite{FRIDEMFUEWIC07}. The family
${\mathcal G}_n$ we use is the set of all squares.

\subsection{An example}
As a simple example we consider the function $F: \rz^2 \rightarrow \rz$
\begin{eqnarray*}
  F(x,y) = 10\exp(-x^2-2y^2)
\end{eqnarray*}
on a $50\times 50$ grid on $[-7,4]^2$ with added normal noise,
$\eps_i\sim N(0,1).$ Figure \ref{fig:tpsdat} shows the function $F$
and its contaminated version together with the thin plate splines
reconstruction using generalized cross-validation and the weighted
smoothing spline method. The main drawback of weighted
thin plate splines is the numerical difficulty of calculating them for
larger grids.
\begin{figure}
  \centering
  \psfig{file=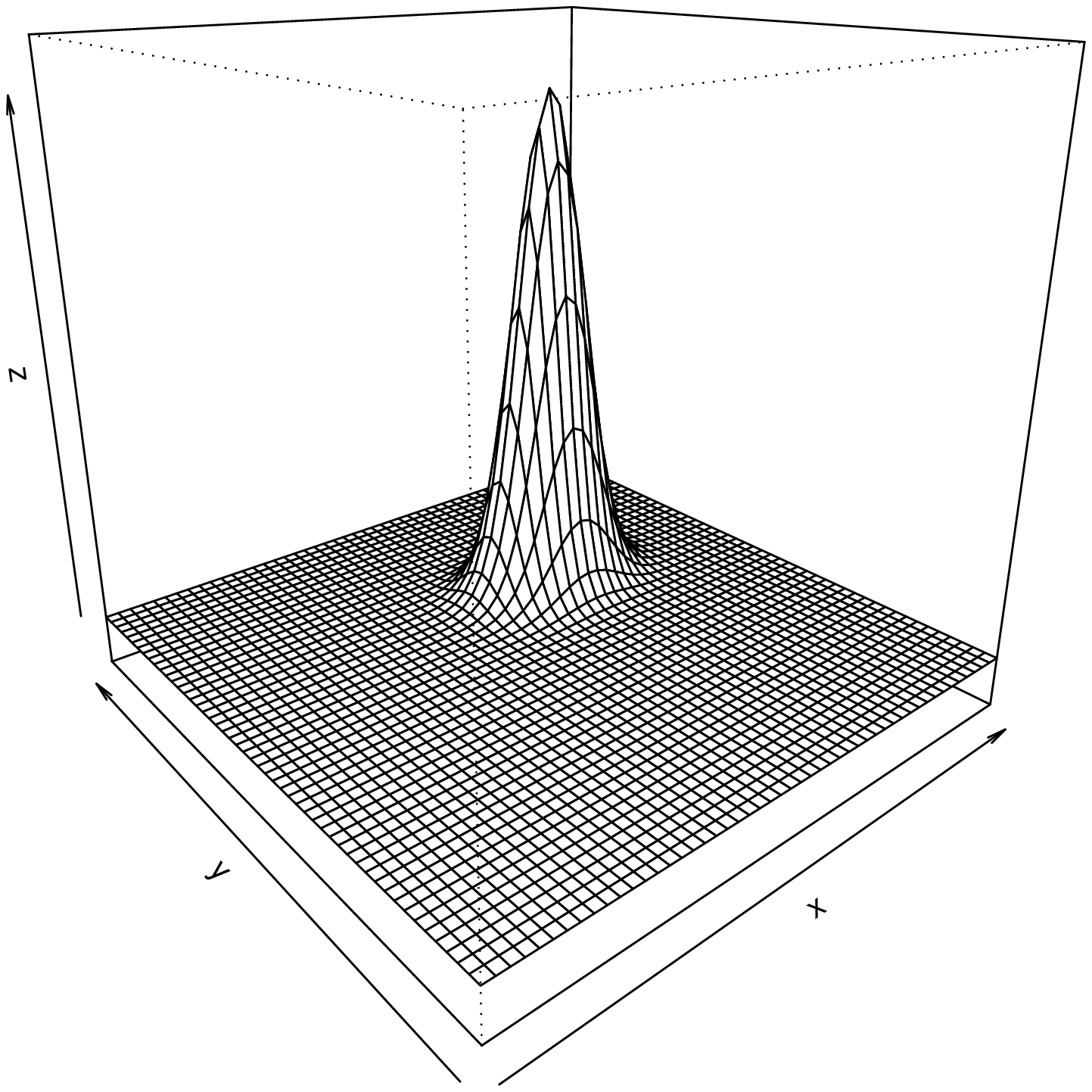,angle=270,width=7cm}
  \psfig{file=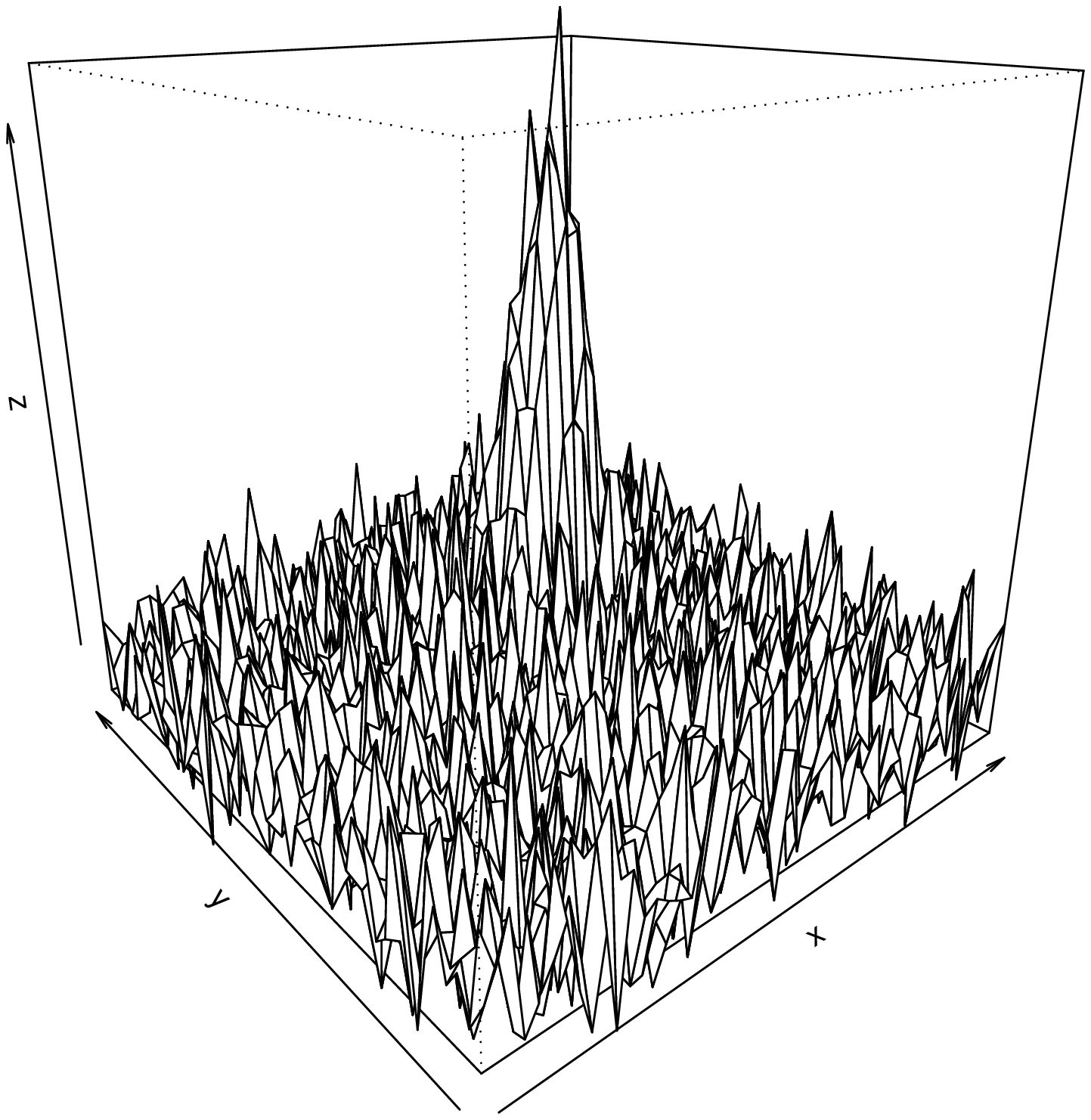,angle=270,width=7cm}\\
  \psfig{file=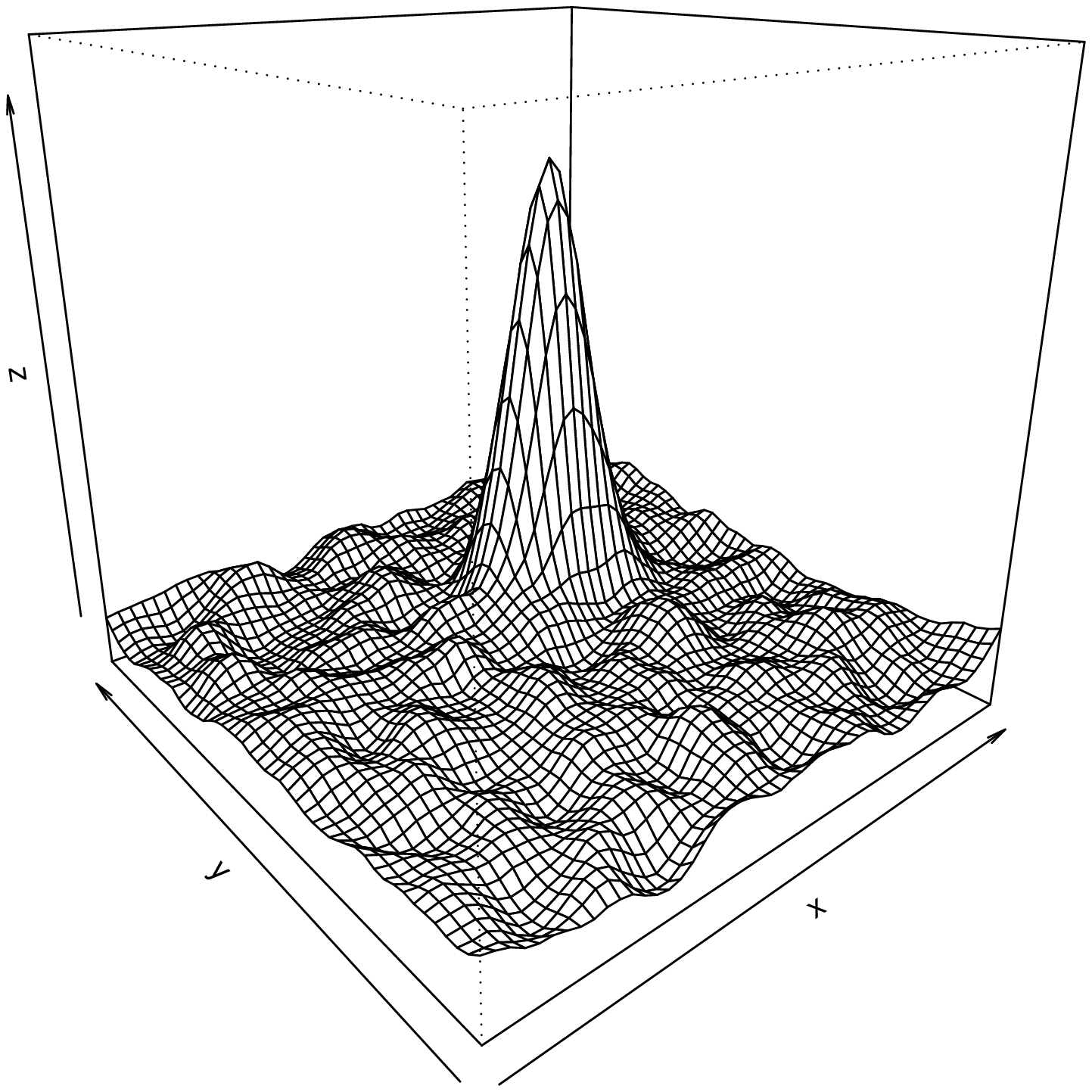,angle=270,width=7cm}
  \psfig{file=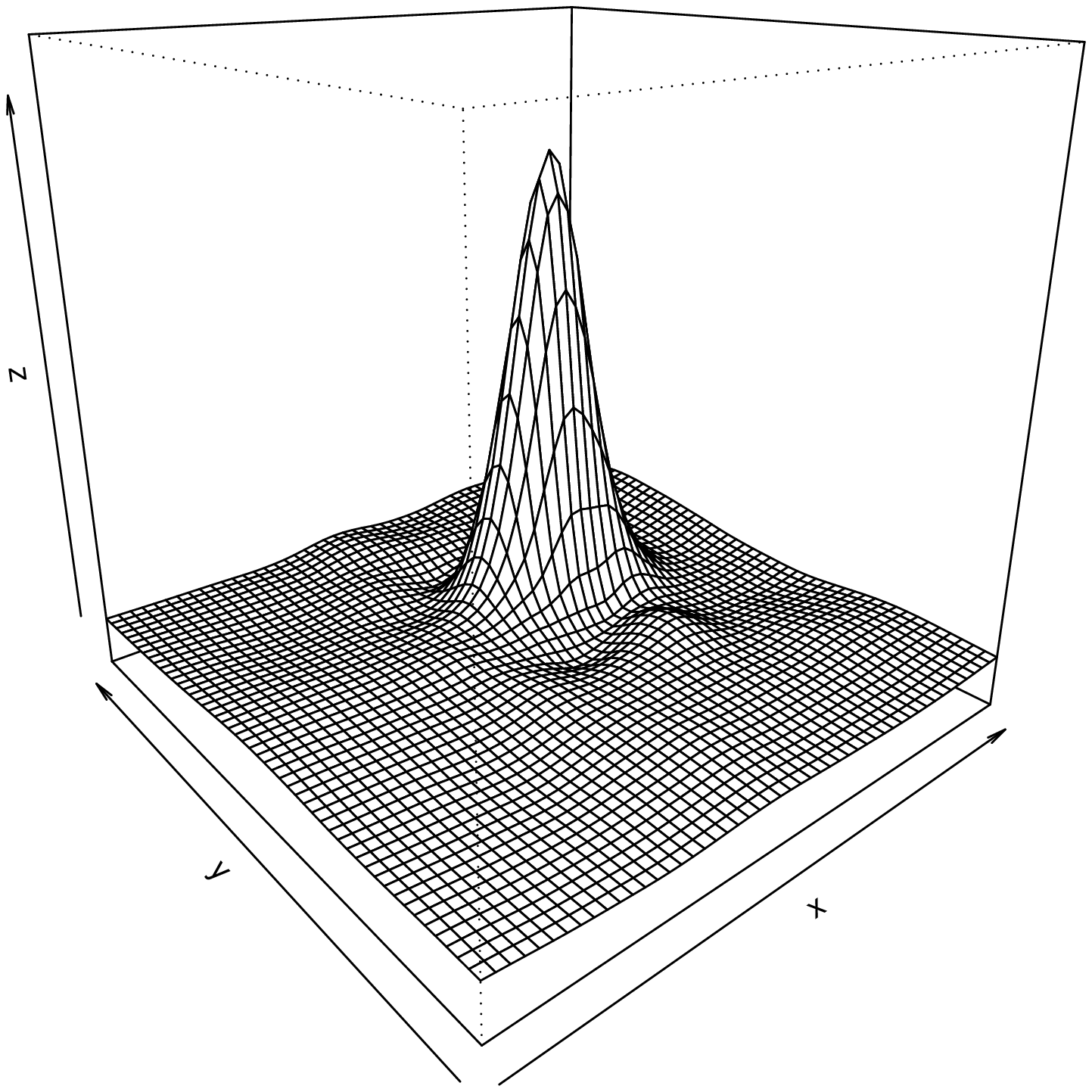,angle=270,width=7cm}
  \caption{{\footnotesize Top row: original function (left) and the
      noisy data (right). Bottom row: thin plate approximation using
      GCV (left) and the automatically weighted version
      (right). \label{fig:tpsdat}}}
\end{figure}

\section{Asymptotics} \label{asymp}
\subsection{Weighted smoothing splines}
Weighted smoothing splines may be seen as a heuristic method for
solving
\begin{eqnarray} \label{minundintg2}
 \min\, R(g) \quad\text{ s.~t.}\quad  g \in {\mathcal
  A}({\mathbold Y}_n, \sigma_n,{\mathcal I}_n,\tau_n).
\end{eqnarray} 
The resulting function $f_n$
is defined by an algorithm and in the absence of a proof that it
yields at least an approximate solution, that is 
\[\int_0^1f_n^{(2)}(t)^2\,dt \le \, K\min_{ g \in {\mathcal
  A}({\mathbold Y}_n, \sigma_n,{\mathcal
  I}_n,\tau_n)}\,\int_0^1g^{(2)}(t)^2\,dt\] 
for some constant $K>0,$ we can either establish a rate of convergence
on this assumption or we can try and analyse the algorithm. In the
first case we are lead to a rate of convergence in the supremum norm
of order $(\log(n)/n)^{3/8}.$ Analysing the algorithm as it
stands would essentially involve proving that it solves the
minimization problem at least approximately. We therefore analyse a
modified version of the procedure. We assume that
the design points are of the form $t_i=i/n$ and that the data are
generated as in (\ref{basicmod}) with 
\begin{eqnarray} 
f^{(1)}(t)-f^{(1)}(0)&= &\int_0^t f^{(2)}(u)\,du,\label{fcond1}\\
\int_0^1f^{(2)}(u)^2\,du &<&\infty.\label{fcond2}
\end{eqnarray}
For a given function $g$ we denote the vector of values of $g$ at
the design points by ${\mathbold g}_n.$ We consider firstly the case
of a global $\lambda$ and denote the solution of (\ref{eq:WSS}) with
$\lambda_1=\ldots=\lambda_n=\lambda$ by ${\mathbold {\tilde f}}_n(\lambda).$ It
can be shown that  ${\mathbold {\tilde f}}_n(\lambda)$ is a solution of 
\begin{eqnarray} \label{eq:SSndim}
  \min\quad
  S_{\lambda}({\mathbold g}_n):=\sum_{i=1}^n\lambda(Y(t_i)-g_n(t_i))^2+
  {\mathbold g}_n^t\Omega_n {\mathbold g}_n
\end{eqnarray}
where $\Omega_n$ is an $n\times n$-non-negative definite matrix with
normalized eigenvectors ${\mathbold e}_{ni}$ and corresponding
eigenvalues $\gamma_{ni}, 1\le i \le n$ with
$\gamma_{n1}=\gamma_{n2}=0.$ The remaining eigenvalues satisfy the
inequalities  
\begin{eqnarray} \label{eigeninequdim}
  c_1\frac{i^4}{n}\, \le\, \gamma_{ni}\, \le c_2\frac{i^4}{n}, \quad 3
\le i\le n
\end{eqnarray}
with the constants $c_1$ and $c_2$ being independent of $n$ (see
\cite{UTRE83}). For an interval
${\tilde J} \subset \{1,\ldots,n\}$ we denote by $\theta_I$ the vector whose
elements $\theta_i$ are $1/\sqrt{\vert{\tilde J}\vert}$ for $i \in
{\tilde J}$ and 0 otherwise. We see that $\Vert \theta_I\Vert=1$ and
for the solution ${\mathbold {\tilde f}}_n(\lambda)$ of (\ref{eq:SSndim})
the $w({\mathbold Y}_n,I,{\tilde f}_n(\lambda))$ of (\ref{defwifn})
are given by 
\begin{eqnarray*} \label{wifnsol}
w({\mathbold Y}_n,I,{\tilde f}_n(\lambda))=\theta_{{\tilde J}(I)}I^t
({\mathbold Y}_n-{\tilde {\mathbold f}}_n(\lambda)),\quad I \in{\mathcal I}_n
\end{eqnarray*}
where ${\tilde J}(I)$ is the interval of $ \{1,\ldots,n\}$ which gives
the indices $i$ with $t_i \in I.$
We have
\begin{theorem}
\label{th0}
\quad
\begin{itemize}
\item[(a)] ${\tilde {\mathbold f}}_n^t(\lambda) \Omega_n {\tilde
    {\mathbold f}}_n(\lambda)$ is an
increasing function of $\lambda.$
\item[(b)] $\ex\left({\tilde {\mathbold f}}_n^t(\lambda) \Omega_n {\tilde
      {\mathbold f}}_n(\lambda)\right) \le cn^{1/4}\lambda^{5/4}$ for some
  constant $c.$ 
\item[(c)] There exists a constant $A>0$ such that for all $\lambda >
  A/\log n$ and for all ${\mathcal I}_n$ with $\vert {\mathcal I}_n\vert \le
  qn$ for some fixed $q$ and for all $\tau > 2$ we have 
\[ \lim_{n \rightarrow \infty}\pr\big( \max_{ I \in {\mathcal I}_n}\vert
w({\mathbold Y}_n,I,{\tilde f}_n(\lambda))\vert \le \sigma \sqrt{\tau
  \log n\,}\,\big) =1.\] 
\end{itemize}
\end{theorem} 
{\bf Proof.} (a) In the following $I_n$ denotes the identity
matrix. The solution ${\tilde {\mathbold f}}_n(\lambda)$ of
(\ref{eq:SSndim}) is given by  
\[{\tilde {\mathbold f}}_n(\lambda)=\lambda (\lambda
I_n+\Omega_n)^{-1}{\mathbold Y}_n\]
and on writing ${\mathbold Y}_n= \sum_{i=1}^n \eta_{ni}{\mathbold
  e}_{ni}$ we obtain 
\[{\tilde {\mathbold f}}_n^t(\lambda) \Omega_n {\tilde
  {\mathbold f}}_n(\lambda)=\lambda^2\sum_{i=3}^n
\frac{\eta_{ni}^2\gamma_{ni}}{(\lambda+\gamma_{ni})^2}\] 
from which the claim follows on noting that $\gamma_{ni} > 0$ for $i
\ge 3.$ \\
\quad\\
(b) We have
\[ {\tilde {\mathbold f}}_n^t(\lambda)\Omega_n {\tilde
  {\mathbold f}}_n(\lambda) =
\lambda^2{\mathbold Y}_n^t(\lambda I_n +\Omega_n)^{-1}\Omega_n(\lambda I_n
+\Omega_n)^{-1}{\mathbold Y}_n\]  
and hence
\begin{eqnarray*}
  \lefteqn{\ex\left({\tilde {\mathbold f}}_n^t(\lambda)\Omega_n {\tilde
    {\mathbold f}}_n(\lambda)\right)}\\
&=&\lambda^2 {\mathbold f}_n^t (\lambda I_n
+\Omega_n)^{-1}\Omega_n(\lambda I_n +\Omega_n)^{-1}{\mathbold f}_n\\&&
+\sigma^2\ex\left(\lambda^2{\mathbold Z}_n^t(\lambda I_n
  +\Omega_n)^{-1}\Omega_n(\lambda I_n +\Omega_n)^{-1}{\mathbold Z}_n\right).
\end{eqnarray*}
Arguing as above we obtain
\begin{eqnarray*}
  \lambda^2 {\mathbold f}_n^t(\lambda I_n
  +\Omega_n)^{-1}\Omega_n(\lambda I_n +\Omega_n)^{-1}{\mathbold f}_n\\
=\lambda^2\sum_3^n\alpha_{ni}^2\frac{\gamma_{ni}}{(\lambda
    +\gamma_{ni})^2} \\
\le \sum_3^n\alpha_{ni}^2\gamma_{ni} = {\mathbold f}_n^t \Omega
{\mathbold f}_n
\end{eqnarray*}
and
\begin{eqnarray*}
 \ex(\lambda^2{\mathbold Z}_n^t(\lambda I_n +\Omega_n)^{-1}\Omega_n(\lambda
  I_n +\Omega_n)^{-1}{\mathbold Z}_n) \\
=\lambda^2\sum_3^n
  \frac{\gamma_{ni}}{(\lambda+\gamma_{ni})^2}.
\end{eqnarray*}
On splitting the last sum into two parts, from $i=3$ to
$i=n^{1/4}\lambda^{1/4}$ and from $i=n^{1/4}\lambda^{1/4}$ to $i=n$
and on using (\ref{eigeninequdim}) we see that
\[\ex(\lambda^2{\mathbold Z}_n^t(\lambda I_n +\Omega_n)^{-1}\Omega_n(\lambda
  I_n +\Omega_n)^{-1}{\mathbold Z}_n) \le c n^{1/4}\lambda^{5/4}\]
for some constant $c.$ 
\quad\\
(c) We have 
\[{\mathbold Y}_n-{\tilde{\mathbold f}}_n(\lambda)= (\lambda I_n
+\Omega_n)^{-1}\Omega_n{\mathbold Y}_n.\] 
and on writing ${\mathbold Y}_n = {\mathbold f}_n + {\mathbold Z}_n$
we obtain
\begin{eqnarray*}
{\mathbold Y}_n-{\tilde {\mathbold f}}_n(\lambda)={\mathbold h}_n+\delta_n
\end{eqnarray*}
with
\begin{eqnarray*}
{\mathbold h}_n&=&(\lambda I_n +\Omega_n)^{-1}\Omega_n{\mathbold f}_n,\\
\delta_n&=&\sigma (\lambda I_n +\Omega_n)^{-1}\Omega_n {\mathbold Z}_n.
\end{eqnarray*}
On writing ${\mathbold f}_n = \sum_1^n \alpha_{ni} {\mathbold e}_{ni}$
we obtain 
\[{\mathbold h}_n= \sum_3^n \alpha_{ni} \frac{\gamma_{ni}}{(\lambda +
    \gamma_{ni})}{\mathbold e}_{ni} \]
and hence
\begin{eqnarray*}
  \Vert {\mathbold h}_n\Vert^2 &=& \sum_3^n\alpha_{ni}^2
\frac{\gamma_{ni}^2}{(\lambda +  \gamma_{ni})^2}\\
&=&\frac{1}{\lambda}\sum_3^n\alpha_{ni}^2
\frac{\gamma_{ni}^2/\lambda}{(1 + \gamma_{ni}/\lambda)^2}
\le\frac{1}{\lambda} \sum_3^n\alpha_{ni}^2\gamma_{ni}. 
\end{eqnarray*}
As ${\mathbold f}_n^t\Omega_n{\mathbold f}_n =
\sum_3^n\alpha_{ni}^2\gamma_{ni}$ we see that at least asymptotically 
\begin{eqnarray*} \label{bndhn}
\Vert{\mathbold h}_n\Vert^2 \le \frac{1}{\lambda}{\mathbold
  f}_n^t\Omega_n{\mathbold f}_n .
\end{eqnarray*}
We turn to $\delta_n.$ We write ${\mathbold Z}_n = \sum_1^n
Z_{ni}^*{\mathbold e}_{ni}$ where, because of the transformation is
orthonormal, the $Z_{ni}^*$ are i.i.d. standard Gaussian random
variables. It follows 
\[\delta_n = \sigma\sum_3^n Z_{ni}^*\frac{\gamma_{ni}}{(\lambda
    +\gamma_{ni})}{\mathbold e}_{ni}\]
and on writing $\theta_I=\sum_1^n\theta_{ni}{\mathbold e}_{ni}$ we obtain 
\[\ex((\theta_I^t\delta_n)^2)=\sigma^2\sum_3^n
\theta_{ni}^2\left(\frac{\gamma_{ni}}{\lambda+\gamma_{ni}}\right)^2
\le \sigma^2.\]
The claim of the theorem follows from the usual upper bound for the
tail of a Gaussian distribution.\hfill $\Box$\\

 We consider the following modified procedure. We consider the
solutions ${\tilde {\mathbold f}}_n(\lambda)$ of (\ref{eq:SSndim}) and
determine the smallest value of $\lambda$ for which ${\tilde
  f}_n(\lambda)\in {\mathcal A}({\mathbold Y}_n,{\mathcal
  I}_n,\tau_n).$ It follows from (c) of Theorem \ref{th0} this 
smallest value is asymptotically with arbitrarily large probability
smaller $A/\log n.$ If we denote this solution by ${\mathbold {\tilde
  f}}_n(\lambda^*_n)$ then (a) and (b) of Theorem \ref{th0} imply
\begin{eqnarray} \label{smthsol}
\lim_{c \rightarrow \infty} \lim_{n\rightarrow \infty}\pr\Big({\tilde
  {\mathbold 
    f}}_n(\lambda^*_n)^t \Omega_n {\tilde {\mathbold
    f}}_n(\lambda^*_n) \le cn^{1/4}(\log n)^{-5/4}\Big)=1. 
\end{eqnarray}
for some $c >0.$ Let ${\tilde {\mathbold f}}_n({\mathbold \lambda})$
be the solution obtained from the weighted smoothing
spline procedures as described in Section \ref{wssproc} respectively. If   
\[ {\tilde {\mathbold f}}_n({\mathbold \lambda})^t \Omega_n {\tilde
  {\mathbold f}}_n({\mathbold \lambda}) \le {\tilde
  {\mathbold f}}_n(\lambda^*_n)^t \Omega_n {\tilde {\mathbold
    f}}_n(\lambda^*_n) \] 
then we accept ${\tilde {\mathbold f}}_n({\mathbold \lambda})$ and
otherwise we accept ${\tilde {\mathbold f}}_n(\lambda^*_n)$  and denote
  the solution by ${\mathbold f}_n^*.$ We have
 
\begin{theorem} \label{th1}
If $f$ satisfies (\ref{fcond1}) and (\ref{fcond2}) and if $\delta_n$
is such that 
\[\lim_{n \rightarrow \infty} \delta_n n^{5/16}(\log
n)^{-9/16}=\infty\]
 then 
\[\sup_{\delta_n \le t \le 1-\delta_n} \vert f_n^*(t)-f(t)\vert=
\text{O}_P((\log n)^{7/32}n^{-11/32}).\] 
\end{theorem}
{\bf Proof.} For a function $g$ satisfying the conditions
(\ref{fcond1}) and (\ref{fcond2}) we  have 
\begin{eqnarray*}
g(t+s)-g(t)&=&\int_0^sg^{(1)}(t+u)\,du=sg^{(1)}(t)+
\int_0^s(g^{(1)}(t+u)-g^{(1)}(t))\,du \\
&=&sg^{(1)}(t)+\int_0^s\left(\int_0^ug^{(2)}(t+v)\,dv\right)\,du
\end{eqnarray*}
and hence
\[\vert g(t+s)-g(t)-sg^{(1)}(t)\vert \le \int_0^s\left(\int_0^u\vert
  g^{(2)}(t+v)\vert\,dv\right)\,du.\]
As
\begin{eqnarray*}
\left(\int_0^u\vert g^{(2)}(t+v)\vert\,dv\right)^2 &=&
\left(\int_0^1\{v\le u\}\vert g^{(2)}(t+v)\vert\,dv\right)^2 \\
&\le & \int_0^1\{v\le u\}^2\,dv
\int_0^1g^{(2)}(t)^2\,dv=u\int_0^1g^{(2)}(t)^2\,dv 
\end{eqnarray*}
for $u$ with $t+u <1$ by Cauchy-Schwarz we obtain
\begin{eqnarray*}
\vert g(t+s)-g(t)-sg^{(1)}(t)\vert &\le&
\int_0^su^{1/2}\left(\int_0^1g^{(2)}(t)^2\,dv\right)^{1/2}\,du\\
&=&\frac{2}{3}s^{3/2}\left(\int_0^1g^{(2)}(t)^2\,dv
\right)^{1/2}.
\end{eqnarray*}
On combining this with the corresponding inequality for $\vert
g(t-s)-g(t)+sg^{(1)}(t)\vert$ we conclude
\begin{equation} \label{inequ1}
 \vert g(t+s)+g(t-s)-2g(t)\vert \le \frac{4}{3} s^{3/2}\Big(\int_0^1
g^{(2)}(u)^2\,du\Big)^{1/2}
\end{equation}
At this point to simplify the proof we assume that ${\mathcal I}_n$ is
the family of all intervals of the form $[t_i,t_j].$ The only effect
of taking ${\mathcal I}_n$ to be the dyadic set of intervals is that
the constants in the estimates below are somewhat larger. Consider now
point $t_j=j/n$ and the interval $I_{j,k}=[t_{j-k},t_{j+k}].$ As
$f^*_n$ lies in ${\mathcal I}_n$ we have
\begin{equation} \label{inequ2}
\frac{1}{\sqrt{2k+1}}\Big\vert \sum_{i=-k}^k
\left(f_n^*\left(\frac{j+i}{n}\right)-f\left(\frac{j+i}{n}\right)\right)
\Big \vert \le \sigma_n\sqrt{\tau_n\log n}.
\end{equation}
We intend to use (\ref{inequ1}) with $g=g_n=f_n^*-f$, $t=j/n$ and
$s=i/n$. Firstly we note that for this $g$ it follows from (\ref{fcond2}) and
(\ref{smthsol})  that
\[\int_0^1g^{(2)}(t)^2\,dv =\text{O}_P\left( n^{1/4}(\log
  n)^{-5/4}\right). \]
From this and (\ref{inequ1}) we deduce
\begin{eqnarray*}
\lefteqn{ f_n^*\left(\frac{j+i}{n}\right) +f_n^*\left(\frac{j-i}{n}\right)
-f\left(\frac{j+i}{n}\right) -f\left(\frac{j-i}{n}\right)=
2\left (f_n^*\left(\frac{j}{n}\right)-
  f\left(\frac{j}{n}\right)\right)} \\
&& \hspace{12cm}+ \,R_n
\end{eqnarray*} 
with
\[R_n= \left(\frac{i}{n}\right)^{3/2}\text{O}_P\left( n^{1/8}(\log
  n)^{-5/8}\right).\]
On using this in (\ref{inequ2}) we obtain after a short calculation
\[\left \vert f_n^*\left(\frac{j}{n}\right)- f\left(\frac{j}{n}\right)
  \right \vert \le \left(\frac{k}{n}\right)^{3/2}\text{O}_P\left(
      n^{1/8}(\log n)^{-5/8}\right)+\sigma_n\sqrt{\frac{\tau_n \log n}{2k}}.\]
As $f$ is continuous it is easy to prove that $\lim_{n \rightarrow
  \infty} \sigma_n=\sigma$ and as, as already noted, $\lim_{n
  \rightarrow \infty} \tau_n=2$ we deduce
\[\left \vert f_n^*\left(\frac{j}{n}\right)- f\left(\frac{j}{n}\right)
  \right \vert \le \text{O}_P\left( \left(\frac{k}{n}\right)^{3/2}
    n^{1/8}(\log n)^{-5/8}+\sqrt{\frac{\log n}{k}}\,\right).\] 
The result follows on choosing $k=n^{11/16}(\log n)^{9/16}$.\hfill
  $\Box$\\

We note that for the solution ${\hat f}_n$ of (\ref{minundintg2}) we
have 
\[\int_0^1 {\hat f}_n^{(2)}(t)^2\,dt \le \int_0^1 f^{(2)}(t)^2\,dt.\]
This means that we can replace the term $\text{O}_P\left( n^{1/8}(\log
  n)^{-5/8}\right)$ above by $\text{O}_P(1).$ The same argument now
leads to the rate of convergence $(\log n/n)^{3/8}$ mentioned
above.

\subsection{Weighted thin plate smoothing splines}
The method of prove can be extended to obtain an analogous result for
weighted thin plate smoothing splines. As the calculations are
somewhat longer we only indicate how to do this. The
estimates  (\ref{eigeninequdim}) are replaced by
\begin{eqnarray*} \label{eigeninequdim2}
  c_1\frac{i^2}{n}\, \le\, \gamma_{ni}\, \le c_2\frac{i^2}{n}, \quad 3
\le i\le n
\end{eqnarray*}
with the constants $c_1$ and $c_2$ being independent of $n$ (see
\cite{UTRE88}). From this the same method of proof used for Theorem
\ref{th0} leads to a corresponding result. The family ${\mathcal I}_n$
is taken to be the family of squares and now a two-dimensional
version of the argument leading to Theorem \ref{th1} gives the result.

\section{Acknowledgments}
We gratefully acknowledge the financial support of the
Sonderforschungsbereich 475, `Komplexit\"atsreduktion in multivariaten
Datenstrukturen', Department of Statistics, University of Dortmund.

We also acknowledge the helpful comments of an anonymous referee and
an Associate Editor which lead to a great improvement in clarity and
presentation.

\renewcommand{\baselinestretch}{1}
\normalsize 
\bibliographystyle{apalike}

\end{document}